\documentclass[final]{aipproc}

\usepackage{amssymb}
\usepackage{amsmath}

\layoutstyle{6x9}

\begin{document}

\newcommand{\half}{\mbox{$\textstyle \frac{1}{2}$}}
\newcommand{\roothalf}{\mbox{$\textstyle \frac{1}{\sqrt{2}}$}}
\newcommand{\re}{{\rm e}}
\newcommand{\ri}{{\rm i}}
\newcommand{\rd}{{\rm d}}
\newcommand{\mba}{\boldsymbol{\alpha}}
\newcommand{\mbb}{\boldsymbol{\beta}}
\newcommand{\mbc}{\boldsymbol{\gamma}}
\newcommand{\mbd}{\boldsymbol{\delta}}
\newcommand{\mbp}{\boldsymbol{\rho}}
\newcommand{\mbq}{\boldsymbol{\sigma}}
\newcommand{\mbr}{\boldsymbol{\tau}}

\title{Twistor cosmology and quantum space-time}

\author{Dorje~C.~Brody}{
  address={Blackett Laboratory, Imperial College,
London SW7 2BZ, UK} }

\author{Lane~P.~Hughston}{
  address={Department of Mathematics, King's College
London, The Strand, London WC2R 2LS, UK} }

\begin{abstract}
The purpose of this paper is to present a model of a `quantum
space-time' in which the global symmetries of space-time are
unified in a coherent manner with the internal symmetries
associated with the state space of quantum-mechanics. If we take
into account the fact that these distinct families of symmetries
should in some sense merge and become essentially
indistinguishable in the unified regime, our framework may provide
an approximate description of or elementary model for the
structure of the universe at early times. The quantum elements
employed in our characterisation of the geometry of space-time
imply that the pseudo-Riemannian structure commonly regarded as an
essential feature in relativistic theories must be dispensed with.
Nevertheless, the causal structure and the physical kinematics of
quantum space-time are shown to persist in a manner that remains
highly analogous to the corresponding features of the classical
theory. In the case of the simplest conformally flat cosmological
models arising in this framework, the twistorial description of
quantum space-time is shown to be effective in characterising the
various physical and geometrical properties of the theory. As an
example, a sixteen-dimensional analogue of the
Friedmann-Robertson-Walker cosmologies is constructed, and its
chronological development is analysed in some detail. More
generally, whenever the dimension of a quantum space-time is an
even perfect square, there exists a canonical way of breaking the
global quantum space-time symmetry so that a generic point of
quantum space-time can be consistently interpreted as a quantum
operator taking values in Minkowski space. In this scenario, the
breakdown of the fundamental symmetry of the theory is due to a
loss of quantum entanglement between space-time and internal
quantum degrees of freedom. It is thus possible to show in a
certain specific sense that the classical space-time description
is an emergent feature arising as a consequence of a quantum
averaging over the internal degrees of freedom. The familiar
probabilistic features of the quantum state, represented by
properties of the density matrix, can then be seen as a by-product
of the causal structure of quantum space-time.
\end{abstract}

\maketitle


\section{Introduction}

This article is concerned with a programme that has as its goal
the development of a theory of quantum space-time. In this
programme, an outline of which will be given in more detail
shortly, an important role is played by certain higher-dimensional
analogues of spinors and twistors. It will be useful to begin,
therefore, by remarking that there are two distinct notions of how
one extends the concept of spinor into higher dimensions. This
fundamental dichotomy arises in association with the fact that in
four-dimensional space-time there is a local isomorphism between
the Lorentz group $SO(1,3)$ and the spin transformation group
$SL(2,{\mathbb C})$. In higher dimensions, however, this relation
breaks down and as a consequence we are left with two concepts of
spinors---one for the group $SO(N,{\mathbb C})$, and one for the
group $SL(r,{\mathbb C})$.

The spinors associated with $SO(N,{\mathbb C})$, where we allow
also for various possible signatures in the quadratic form
defining these orthogonal or pseudo-orthogonal transformations
when we specialise to the real subgroup $SO(p,q)$ with $p+q=N$,
are the so-called Cartan spinors~\cite{Budinich,Cartan,
Chevalley,Penrose9}. The study of Cartan spinors has a long and
interesting history, and there is a beautiful geometry associated
with these spinors. There are also various specific cases of great
interest---for example, the Cartan spinors associated with the
group $SO(2,4)$ are Penrose's twistors; and the Cartan spinors
associated with $SO(8)$ are intimately linked with the Cayley
numbers (octonions) and the exceptional Lie groups. There are also
a number of interesting connections between Cartan spinors and
massless fields in higher dimensions~\cite{Hughston4,
Hughston6,Hughston5}.

The spinors associated with $SL(r,{\mathbb C})$, which are usually
now called `hyperspinors', have the advantage of being more
directly linked with quantum mechanics. In fact, we shall show
later that a naturally relativistic model for hyperspin arises
when one considers `multiplets' of two-component spinors, i.e.
expressions of the form $\xi^{{\bf A}i}$ and $\eta^{{\bf
A}^\prime}_i$, where ${\bf A},{\bf A}'$ are standard spinor
indices, and $i=1,2, \ldots,n$ is an `internal' index. In the
general case ($n=\infty$) we then think of $\xi^{{\bf A}i}$ as an
element of the tensor product space ${\mathbb S}^{{\bf A}i}
={\mathbb S}^{\bf A}\otimes{\mathbb H}^i$, where ${\mathbb S}^{\bf
A}$ is the complex vector space of two-component spinors, and
${\mathbb H}^i$ is an infinite-dimensional complex Hilbert space.

There is also a link, arising through a further extension of this
idea, between hyperspinor theory and the theory of multi-twistor
(hypertwistor) systems. Indeed, we find that the theory of
hyperspin constitutes a natural starting place for building up a
theory of quantum geometry or, as we shall call it here,
\emph{quantum space-time}. In summary, we shall be taking the
left-hand path in the following diagram:

\vspace{0.2cm}
\begin{picture}(300,100)(0,0)
\thicklines \put(155,80){$SL(2,{\mathbb C})\sim SO(1,3)$}
\put(125,8){$SL(r,{\mathbb C})$} \put(240,8){$SO(N,{\mathbb C})$}
\put(105,-5){Hyperspinors\ $\checkmark$} \put(225,-5){Cartan
spinors} \put(196,72){\vector(-1,-1){50}}
\put(204,72){\vector(1,-1){50}}
\end{picture}
\vspace{0.4cm}

The hyperspinor route has the virtue that the resulting space-time
has a rich causal structure associated with it, and as a
consequence is unusually well-positioned to form the geometrical
basis of a physical theory.

\section{Relativistic causality}

To start, let us review briefly the role of two-component spinors
in the description of four-dimensional Minkowskian space-time
geometry. In what follows we use bold upright Roman letters to
denote two-component spinor indices, and we adopt the standard
conventions for the algebra of two-component
spinors~\cite{Penrose2,Penrose8,Penrose9,Pirani}. Then we have the
following correspondence between two-by-two Hermitian matrices and
space-time points, relative to some origin:
\begin{eqnarray}
x^{{\bf AA}^\prime}\quad  ({\bf A},{\bf A}^\prime=1,2)\qquad
\longleftrightarrow \qquad x^{\rm a} \quad ({\rm a}=0,1,2,3).
\end{eqnarray}
More explicitly, in a standard basis this correspondence is given
by
\begin{eqnarray}
\frac{1}{\sqrt{2}} \left( \begin{array}{ll} t+z & x+\ri y \\ x-\ri
y & t-z \end{array} \right) \quad \longleftrightarrow \quad
(t,x,y,z). \label{eq:2}
\end{eqnarray}
We then have the fundamental relation
\begin{eqnarray}
2\det(x^{{\bf AA}^\prime}) =  t^2-x^2-y^2-z^2 , \label{eq:3}
\end{eqnarray}
from which it follows that two-component spinors are connected
both with quantum mechanics and with the causal structure of
space-time. It is a peculiar aspect of relativistic physics that
there is this link between (a) the spin degrees of freedom of spin
one-half particles, and (b) the causal geometry of
four-dimensional space-time.

Let us pursue this idea now in a little more detail, and then
extend it to higher dimensions. For the interval $r^{{\bf
AA}^\prime}$ between a pair of points $x^{{\bf AA}^\prime}$ and
$y^{{\bf AA}^\prime}$ in Minkowski space-time we write
\begin{eqnarray}
r^{{\bf AA}^\prime} = x^{{\bf AA}^\prime}-y^{{\bf AA}^\prime},
\end{eqnarray}
from which it follows that
\begin{eqnarray}
2\det(r^{{\bf AA}^\prime}) = \epsilon_{\bf AB}\epsilon_{{\bf
A}^\prime{\bf B}^\prime} r^{{\bf AA}^\prime} r^{{\bf BB}^\prime},
\end{eqnarray}
where $\epsilon_{\bf AB}$ is the antisymmetric spinor. Hence if we
adopt the standard `index clumping' convention and write ${\rm
a}={\bf AA}'$, ${\rm b}={\bf BB}'$, and so on, according to which
a pair of spinor indices, one primed and the other unprimed,
corresponds to a lower case space-time vector index, then we can
write
\begin{eqnarray}
\epsilon_{\bf AB}\epsilon_{{\bf A}^\prime{\bf B}^\prime} r^{{\bf
AA}^\prime} r^{{\bf BB}^\prime} = g_{\rm ab} r^{\rm a} r^{\rm b}
\end{eqnarray}
for the corresponding squared space-time interval, and thus we are
able to identity
\begin{eqnarray}
g_{\rm ab} = \epsilon_{\bf AB}\epsilon_{{\bf A}^\prime{\bf
B}^\prime} \label{eq:7}
\end{eqnarray}
as the metric of Minkowski space.

There are essentially three different situations that can arise
for the interval $r^{\rm a}$, each of which represents a certain
level of degeneracy. The first case is $g_{\rm ab}r^{\rm b}=0$;
the second case is $g_{\rm ab}r^{\rm b}\neq0$ and $g_{\rm
ab}r^{\rm a}r^{\rm b}=0$; and the third case is $g_{\rm ab}r^{\rm
a}r^{\rm b}\neq0$. Each of these cases gives rise to a canonical
form for the interval $r^{{\bf AA}^\prime}$, with various
sub-cases, which can be summarised as follows:
\begin{itemize}
\item[(i)] $g_{\rm ab}r^{\rm b}=0$:
\begin{eqnarray}
\begin{array}{ll} r^{{\bf AA}^\prime}=0 & \textrm{zero
separation}
\end{array} \nonumber
\end{eqnarray}
\item[(ii)] $g_{ab} r^a r^b=0$:
\begin{eqnarray}
\begin{array}{ll} r^{{\bf AA}^\prime}=\alpha^{\bf A}
{\bar\alpha}^{{\bf A}^\prime} & \textrm{future-pointing null
separation} \\ r^{{\bf AA}^\prime}=-\alpha^{\bf A}
{\bar\alpha}^{{\bf A}^\prime} & \textrm{past-pointing null
separation} \end{array}  \nonumber
\end{eqnarray}
\item[(iii)] $g_{ab} r^a r^b\neq0$:
\begin{eqnarray}
\begin{array}{ll} r^{{\bf AA}^\prime}=\alpha^{\bf A}
{\bar\alpha}^{{\bf A}^\prime} + \beta^{\bf A}{\bar\beta}^{{\bf
A}^\prime} & \textrm{future-pointing time-like separation} \\
r^{{\bf AA}^\prime} = \alpha^{\bf A}{\bar\alpha}^{{\bf A}^\prime}
- \beta^{\bf A}{\bar\beta}^{{\bf A}^\prime} & \textrm{space-like
separation} \\ r^{{\bf AA}^\prime}=-\alpha^{\bf
A}{\bar\alpha}^{{\bf A}^\prime} - \beta^{\bf A}{\bar\beta}^{{\bf
A}^\prime} & \textrm{past-pointing time-like separation}
\end{array}
\nonumber
\end{eqnarray}
\end{itemize}

\begin{figure}
  \includegraphics[height=.3\textheight,angle=270]{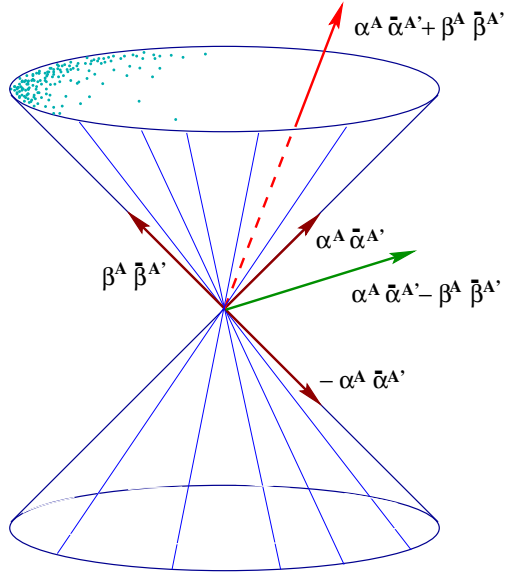}
  \caption{\label{fig:1}\textsl{Causal structure of
  four-dimensional space-time}.
  The canonical form of the spinor decomposition of the
  four-vector associated with a space-time interval determines its
  causal properties. }
\end{figure}

Here it is understood that in case (iii) the spinors $\alpha^{\bf
A}$ and $\beta^{\bf A}$ do not coincide in direction. It is
interesting to note that once the canonical form for $r^{{\bf
AA}^\prime}$ is specified, then so is the causal relationship that
it determines on space-time. This correspondence is illustrated in
Figure~\ref{fig:1}. On the other hand, the specification of
$r^{{\bf AA}^\prime}$ does not completely determine the spinors
$\alpha^{\bf A}$ and $\beta^{\bf A}$. In general there is some
freedom, and this is expressed by a group of transformations. In
particular, if $r^{{\bf AA}^\prime}$ is null, then this freedom is
the phase shift $\alpha^{\bf A}\to {\re}^{\ri\theta} \alpha^{\bf
A}$, and the relevant group is $U(1)$. If $r^{{\bf AA}^\prime}$ is
time-like, then the group is $U(2)$, and if $r^{{\bf AA}^\prime}$
is space-like, the group is $U(1,1)$.

\section{Hyperspin spaces}

The terminology `hyperspinor' is due to
Finkelstein~\cite{Finkelstein1}. Essentially the same concept
(although introduced for different purposes) is also touched on in
\cite{Hughston1}. The idea of a hyperspinor is a simple one---we
replace the two-component spinors associated with four-dimensional
space-time with $r$-component spinors. Thus we can regard
hyperspin space as the vector space ${\mathbb C}^r$ with some
extra structure. In particular, in addition to the original
hyperspin space we have three other vector spaces---the dual
hyperspin space, the complex conjugate hyperspin space, and the
dual complex conjugate hyperspin space.

The theory of hyperspin has been pursued by a number of
authors~\cite{Borowiec1,Borowiec2,Brody2,Finkelstein2,Finkelstein3,
Finkelstein4, Holm1,Holm2,Holm3,Holm4,Solovyov,Urbantke}, and the
material we describe here builds on various aspects of this work.
Let us write ${\mathbb S}^{A}$ and ${\mathbb S}^{A'}$,
respectively, for the complex $r$-dimensional vector spaces of
unprimed and primed hyperspinors. For hyperspinors we use italic
indices to distinguish them from the boldface indices used
exclusively for two-component spinors. It is assumed that
${\mathbb S}^{A}$ and ${\mathbb S}^{A'}$ are related by an
anti-linear isomorphism under the operation of complex
conjugation. Thus if $\alpha^{A}\in{\mathbb S}^{A}$, then under
complex conjugation we have $\alpha^{A}\to {\bar\alpha}^{A'}$,
where ${\bar\alpha}^{A'} \in {\mathbb S}^{A'}$. The dual spaces
associated with ${\mathbb S}^{A}$ and ${\mathbb S}^{A'}$ are
denoted ${\mathbb S}_{A}$ and ${\mathbb S}_{A'}$, respectively. If
$\alpha^{A}\in{\mathbb S}^{A}$ and $\beta_{A}\in{\mathbb S}_{A}$,
then their inner product is denoted $\alpha^{A}\beta_{A}$.
Likewise if $\gamma^{\,A'}\in {\mathbb S}^{A'}$ and
$\delta_{A'}\in {\mathbb S}_{A'}$ then their inner product is
$\gamma^{\,A'}\delta_{A'}$.

We also introduce the totally antisymmetric hyperspinors of rank
$r$ associated with the spaces ${\mathbb S}^{A}$, ${\mathbb
S}_{A}$, ${\mathbb S}^{A'}$, and ${\mathbb S}_{A'}$. These will be
denoted $\varepsilon^{AB\cdots C}$, $\varepsilon_{AB\cdots C}$,
$\varepsilon^{A'B'\cdots C'}$, and $\varepsilon_{A'B'\cdots C'}$,
respectively. The choice of these antisymmetric hyperspinors is
unique up to an overall scale factor. Once a choice has been made
for $\varepsilon_{AB\cdots C}$, then the other epsilon
hyperspinors are determined by the relations
\begin{eqnarray}
\varepsilon^{AB\cdots C} \varepsilon_{AB\cdots C}=r!,\quad
\varepsilon^{A'B'\cdots C'} \varepsilon_{A'B'\cdots C'}=r!,\quad
{\rm and}\quad \varepsilon_{A'B'\cdots C'}=
{\bar\varepsilon}_{A'B'\cdots C'},
\end{eqnarray}
where ${\bar\varepsilon}_{A'B' \cdots C'}$ is the complex
conjugate of $\varepsilon_{AB\cdots C}$.

Now let $\Lambda^m=\Lambda^{A_1A_2\cdots A_m}$ denote the skew
tensor product space ${\mathbb S}^{A_1}\wedge{\mathbb
S}^{A_2}\wedge \cdots\wedge{\mathbb S}^{A_m}$. Using an analogous
notation, we introduce the spaces $\Lambda_m$, $\Lambda^{m'}$, and
$\Lambda_{m'}$ for each $m=0,1,\ldots,r$. Then once the epsilon
hyperspinors have been fixed we have a collection of maps of the
form
\[
\varepsilon_r\!:\,\Lambda^m\to\Lambda_{r-m},\quad\!
\varepsilon^r\!:\,\Lambda_m\to\Lambda^{r-m}, \quad\!
\varepsilon_{r'} \!:\,\Lambda^{m'}\to\Lambda_{r'-m'},\quad\! {\rm
and} \quad\! \varepsilon^{r'}\!: \,
\Lambda_{m'}\to\Lambda^{r'-m'}.
\]
As a consequence, a wide range of algebraic theorems can be
formulated, which are useful in calculations. For example, if
$\alpha^{A} \beta_{A}=0$ then $\beta_{A}$ must be of the form
\begin{eqnarray}
\beta_{A}= \varepsilon_{ABC\cdots D}\alpha^{B} \mu^{C\cdots D}
\end{eqnarray}
for some $\mu^{C\cdots D}\in\Lambda^{r-2}$.

\section{Quantum space-time}

Now we introduce the complex matrix space ${\mathbb C}^{AA'}
={\mathbb S}^{A}\otimes{\mathbb S}^{A'}$. An element
$x^{\,AA'}\in{\mathbb C}^{AA'}$ is said to be \emph{real} if it
satisfies the (weak) Hermitian property $x^{\,AA'}= {\bar
x}^{A'A}$, where ${\bar x}^{A'A}$ is the complex conjugate of
$x^{\,AA'}$. We shall have more to say about weak versus strong
Hermiticity conditions in relation to the idea of symmetry
breaking. We denote the vector space of real elements of ${\mathbb
C}^{AA'}$ by ${\mathbb R}^{AA'}$. The elements of ${\mathbb
R}^{AA'}$ constitute what we call the real quantum space-time
${\mathcal H}^{r^2}$ of dimension $r^2$. We then regard
$\mathcal{C\!H}^{r^2}={\mathbb C}^{AA'}$ as the complexification
of ${\mathcal H}^{r^2}$. Many problems in ${\mathcal H}^{r^2}$ are
best first approached as problems in $\mathcal{C\!H}^{r^2}$, and
hence sometimes although we refer to ${\mathcal H}^{r^2}$ our
operations are actually carried out in $\mathcal{C\!H}^{r^2}$.

Let $x^{\,AA'}$ and $y^{\,AA'}$ be points in ${\mathcal H}^{r^2}$,
and write $r^{\,AA'}=x^{\,AA'} -y^{\,AA'}$ for the corresponding
separation vector, which is independent of the choice of origin.
Using the index-clumping convention we set $x^{\rm a}=x^{\,AA'}$,
$y^{\rm a}=y^{\,AA'}$, $r^{\rm a} = r^{\,AA'}$, and for the
separation of $x^{\rm a}$ and $y^{\rm a}$ in ${\mathcal H}^{r^2}$
we write $r^{\rm a}=x^{\rm a}- y^{\rm a}$. There is a natural
causal structure induced on such intervals by the so-called
`chronometric tensor'. Making use of the index-clumping
convention, we define this fundamental tensor (introduced by
Finkelstein~\cite{Finkelstein1}) by the following basic relation:
\begin{eqnarray}
g_{{\rm ab\cdots c}} = \varepsilon_{AB\cdots C}\,
\varepsilon_{A'B'\cdots C'}.
\end{eqnarray}
The chronometric tensor, which is of rank $r$, is totally
symmetric and is nondegenerate in the sense that for any vector
$r^{\rm a}$ the condition $r^{\rm a}g_{\rm ab\cdots c}=0$ implies
$r^{\rm a}=0$. We say that $x^{\rm a}$ and $y^{\rm a}$ in
${\mathcal H}^{r^2}$ have a `degenerate' separation if the
chronometric form
\begin{eqnarray}
\Delta(r)=g_{{\rm ab\cdots c}}r^{\rm a}r^{\rm b}\cdots r^{\rm c}
\label{eq:10}
\end{eqnarray}
vanishes for $r^{\rm a}=x^{\rm a}- y^{\rm a}$. Degenerate
separation is equivalent to the vanishing of the determinant of
the matrix $r^{\,AA'}$, that is,
\begin{eqnarray}
\varepsilon_{AB\cdots C}\, \varepsilon_{A'B'\cdots C'}
r^{\,AA'}r^{\,BB'}\cdots r^{\,CC'}=0.
\end{eqnarray}

If the hyperspin space has dimension $r=2$, this reduces to the
usual condition for $x^{\rm a}$ and $y^{\rm a}$ to be
null-separated in Minkowski space. For $r>2$, however, the
situation is more complicated since there are various degrees of
degeneracy that can arise between two points of quantum
space-time, of which `nullness' (in the Minkowskian sense) is only
the most extreme.

As an example, consider $r=3$. In this case the quantum space-time
has dimension nine, and the chronometric form is given by
\begin{eqnarray}
\Delta=g_{\rm abc}r^{\rm a}r^{\rm b}r^{\rm c}.
\end{eqnarray}
The different possibilities that can arise for the separation
vector are as follows:
\begin{itemize}
\item[(i)] $g_{\rm abc}r^{\rm c}=0$:
\begin{eqnarray}
\begin{array}{ll} r^{\,AA'}=0 & \textrm{zero separation}
\end{array} \nonumber
\end{eqnarray}
\item[(ii)] $g_{\rm abc}r^{\rm b}r^{\rm c}=0$ and $g_{\rm abc}
r^{\rm c}\neq0$:
\begin{eqnarray}
\begin{array}{ll} r^{\,AA'}=\alpha^A{\bar\alpha}^{A'} &
\textrm{future-pointing null separation} \\ r^{\,AA'}=-\alpha^A
{\bar\alpha}^{A'} & \textrm{past-pointing null
separation}\end{array}  \nonumber
\end{eqnarray}
\item[(iii)] $\Delta=0$ and $g_{\rm abc}r^{\rm b}r^{\rm c} \neq
0$:
\begin{eqnarray}
\begin{array}{ll} r^{\,AA'}=\alpha^A{\bar\alpha}^{A'} +
\beta^A{\bar\beta}^{A'} & \textrm{degenerate time-like
future-pointing separation} \\ r^{\,AA'} =
\alpha^A{\bar\alpha}^{A'} - \beta^A {\bar\beta}^{A'} &
\textrm{degenerate space-like separation} \\
r^{\,AA'}=-\alpha^A{\bar\alpha}^{A'} - \beta^A{\bar\beta}^{A'} &
\textrm{degenerate time-like past-pointing separation}
\end{array}
\nonumber
\end{eqnarray}
\item[(iv)] $\Delta\neq0$ and $g_{\rm ab} r^{\rm a}r^{\rm
b}\neq0$:
\begin{eqnarray}
\begin{array}{ll} r^{\,AA'}=\alpha^A{\bar\alpha}^{A'} +
\beta^A{\bar\beta}^{A'} + \gamma^{\,A}{\bar\gamma}^{\,A'} &
\textrm{future pointing time-like separation} \\ r^{\,AA'} =
\alpha^A {\bar\alpha}^{A'} + \beta^A {\bar\beta}^{A'} -
\gamma^{\,A}
{\bar\gamma}^{\,A'} & \textrm{semi-space-like separation} \\
r^{\,AA'}=\alpha^A {\bar\alpha}^{A'} - \beta^A{\bar\beta}^{A'}-
\gamma^{\,A} {\bar\gamma}^{\,A'} & \textrm{semi-space-like separation}\\
r^{\,AA'}=-\alpha^A {\bar\alpha}^{A'} - \beta^A{\bar\beta}^{A'}-
\gamma^{\,A} {\bar\gamma}^{\,A'} & \textrm{past pointing time-like
separation}
\end{array}
\nonumber
\end{eqnarray}
\end{itemize}

When the separation of two points of quantum space-time is
degenerate, we define the `degree' of degeneracy by the rank of
the matrix $r^{\,AA'}$. Null separation is the case for which the
degeneracy is of the first degree, i.e. where $r^{\,AA'}$ is of
rank one, and thus satisfies a system of quadratic relations of
the following form:
\begin{eqnarray}
\varepsilon_{AB\cdots C}\, \varepsilon_{A'B'\cdots C'}
r^{\,AA'}r^{BB'}=0,
\end{eqnarray}
or equivalently
\begin{eqnarray}
g_{{\rm ab\cdots c}}r^{\rm a}r^{\rm b}=0.
\end{eqnarray}
This implies that $r^{\,AA'}$ can be expressed in the `null' form
\begin{eqnarray}
r^{\,AA'}=\pm \alpha^{A}{\bar\alpha}^{A'}
\end{eqnarray}
for some $\alpha^A$. In the case of degeneracy of the second
degree, $r^{\,AA'}$ is of rank two and satisfies a set of cubic
relations given by
\begin{eqnarray}
\varepsilon_{ABC\cdots D}\, \varepsilon_{A'B'C'\cdots D'}
r^{\,AA'}r^{BB'}r^{CC'}=0,
\end{eqnarray}
or equivalently
\begin{eqnarray}
g_{{\rm abc\cdots d}} r^{\rm a} r^{\rm b}r^{\rm c}=0.
\end{eqnarray}
In this situation $r^{\,AA'}$ can be put into one of the following
three canonical forms:
\begin{itemize}
\item[(a)] $r^{\,AA'}=\alpha^{A} {\bar\alpha}^{A'}+
\beta^A{\bar\beta}^{A'}$,
\item[(b)] $r^{\,AA'}= \alpha^A{\bar\alpha}^{A'}- \beta^{A}
{\bar\beta}^{A'}$,
\item[(c)] $r^{\,AA'}= -\alpha^{A}{\bar\alpha}^{A'}- \beta^{A}
{\bar\beta}^{A'}$.
\end{itemize}
In case (a), $x^{\rm a}$ lies to the future of $y^{\rm a}$, and
$r^{\rm a}$ can be thought of as a degenerate future-pointing
time-like vector. In case (b), $r^{\rm a}$ can be thought of as a
degenerate space-like separation. In case (c), $x^{\rm a}$ lies to
the past of $y^{\rm a}$, and $r^{\rm a}$ is a degenerate
past-pointing time-like vector. A similar analysis can be applied
to degenerate separations of other `intermediate' degrees.

If the determinant of the $r$-by-$r$ weakly Hermitian matrix
$r^{\,AA'}$ is nonvanishing, and $r^{\,AA'}$ is thus of maximal
rank, then the chronometric form $\Delta$ is nonvanishing. In that
case the matrix $r^{\,AA'}$ can be represented in the following
canonical form:
\begin{eqnarray}
r^{\,AA'}=\pm \alpha^{A}{\bar\alpha}^{A'} \pm \beta^{A}
{\bar\beta}^{A'} \pm \cdots \pm \gamma^{\,A}{\bar\gamma}^{\,A'},
\end{eqnarray}
with the presence of $r$ nonvanishing terms, where the $r$
hyperspinors $\alpha^{A},\beta^{A}, \cdots, \gamma^{\,A}$ are all
linearly independent.

Let us write $(p,q)$ for the numbers of plus and minus signs
appearing in the canonical form for the matrix $r^{\,AA'}$ given
above. We call $(p,q)$ the `signature' of $r^{\,AA'}$. The
hyperspinors $\alpha^{A}, \beta^{A}, \cdots, \gamma^{\,A}$ are
determined by the specification of $r^{\,AA'}$ only up to an
overall unitary (or pseudo-unitary) transformation of the form
\begin{eqnarray}
\alpha^{A}_n\to U^m_n \alpha^{A}_m,
\end{eqnarray}
where $n,m=1,2,\ldots,r$, and
\begin{eqnarray}
\alpha^{A}_n = \{\alpha^{A}, \beta^{A},\cdots, \gamma^{\,A}\}.
\end{eqnarray}
The signature $(p,q)$ is nevertheless an invariant of $r^{\,AA'}$.
In the cases for which the signature is $(r,0)$ or $(0,r)$ we say
that $r^{\,AA'}$ is future-pointing time-like or past-pointing
time-like, respectively. Then recalling the definition
(\ref{eq:10}) for the associated chronometric form, we define the
`proper time interval' between the events $x^{\rm a}$ and $y^{\rm
a}$ by the formula
\begin{eqnarray}
\|x-y\| = |\Delta|^{\frac{1}{r}}. \label{eq:3.9}
\end{eqnarray}
In the case $r=2$ we then recover the standard Minkowskian
proper-time interval between the given events.

A remarkable feature of the causal structure of quantum space-time
is that the essential physical features of the causal structure of
Minkowski space are preserved. In particular, the space of
future-pointing time-like vectors forms a convex cone. The same is
true when we consider the structure of the associated momentum
space, from which it follows that we can also give a good
definition of what is meant by `positive energy'.

\section{Equations of motion}

Now suppose that $\lambda\mapsto x^{\,AA'}(\lambda)$ defines a
smooth curve $\gamma$ in ${\mathcal H}^{r^2}$ for $\lambda \in
[a,b]\subset{\mathbb R}$. Then $\gamma$ will be said to be
time-like if the tangent vector along $\gamma$,
\begin{eqnarray}
v^{AA'}(\lambda) = \frac{\rd} {\rd\lambda} x^{\,AA'}(\lambda),
\end{eqnarray}
is time-like and future-pointing. In that case we define the
proper time $s$ elapsed along $\gamma$ by the integral
\begin{eqnarray}
s = \int_a^b \left[ g_{{\rm ab}\cdots{\rm c}} v^{{\rm a}} v^{\rm
b}\cdots v^{\rm c} \right]^{\frac{1}{r}} \rd \lambda .
\label{eq:21}
\end{eqnarray}
In the case of a very small interval, we can also write this in
the `pseudo-Finslerian' form
\begin{eqnarray}
(\rd s)^r = g_{{\rm ab}\cdots{\rm c}}\rd x^{\rm a} \rd x^{\rm b}
\cdots \rd x^{\rm c}.
\end{eqnarray}
In the case $r=2$ this clearly reduces to the standard
pseudo-Riemannian expression for the line element.

\vspace{0.2cm}
\begin{picture}(300,100)
\thicklines \qbezier(110,0)(150,80)(220,68)
\qbezier(220,68)(290,60)(300,80) \put(128,29){\circle*{4}}
\put(255,66){\circle*{4}} \put(132,22){$\lambda=a$}
\put(245,52){$\lambda=b$} \put(180,70){$\gamma$}
\put(118,2){$x^{\rm a}(\lambda)$}
\end{picture}
\vspace{0.4cm}

Now let us consider the condition $\gamma$ must satisfy in order
to be a geodesic in ${\mathcal H}^{r^2}$. Since the geometry is
not Riemannian, the answer may not be entirely obvious. In the
case of a time-like curve, we can choose the proper time as the
parameter along the curve, in which case the resulting affine
parameterisation of the curve is determined up to transformations
of the form $s\to s+k$ where $k$ is a constant. The equation of
motion for the situation in which $\gamma$ is a time-like geodesic
is obtained by an application of the calculus of variations to
formula (\ref{eq:21}). As usual, we assume the variation vanishes
at the endpoints. Writing
\begin{eqnarray}
L=\left( g_{{\rm abc}\cdots{\rm d}} v^{\rm a}v^{\rm b} v^{\rm c}
\cdots v^{\rm d}\right)^{\frac{1}{r}},
\end{eqnarray}
a standard argument shows that $x^{\rm a}(\lambda)$ describes a
geodesic only if the velocity vector $v^{\rm a}$ satisfies the
Euler-Lagrange equation
\begin{eqnarray}
\frac{\rd}{\rd\lambda} \left( \frac{\partial L}{\partial v^{\rm
a}} \right) = 0.
\end{eqnarray}
A calculation shows that this condition is given more explicitly
by
\begin{eqnarray}
g_{{\rm abc}\cdots{\rm d}} \frac{\rd v^{\rm b}} {\rd \lambda}
v^{\rm c} \cdots v^{\rm d} = \phi\,g_{{\rm abc}\cdots{\rm d}}
v^{\rm b} v^{\rm c} \cdots v^{\rm d},
\end{eqnarray}
where
\begin{eqnarray}
\phi=\frac{1}{L}\frac{\rd L}{\rd\lambda}.
\end{eqnarray}
If $\lambda$ is chosen to be proper time, then $\phi=0$ and the
geodesic equation takes to form
\begin{eqnarray}
g_{{\rm abc}\cdots{\rm d}} {\dot v}^{\rm b} v^{\rm c} \cdots
v^{\rm d} = 0, \label{eq:26}
\end{eqnarray}
where the dot denotes differentiation with respect to proper time.

In the case $r=2$ the geodesic equation (\ref{eq:26}) reduces to
the familiar relation ${\dot v}^{\rm a}=0$. We shall now show that
the geodesic equation also implies ${\dot v}^{\rm a}=0$ in the
case of a general quantum space-time with $r>2$. It suffices to
examine the case $r=3$, which will indicate the relevant line of
argument. For $r=3$ the geodesic equation takes the form
\begin{eqnarray}
g_{{\rm abc}} {\dot v}^{\rm b} v^{\rm c}=0,
\end{eqnarray}
which can be expressed in terms of hyperspinors in the form
\begin{eqnarray}
\varepsilon_{ABC}\,\varepsilon_{A'B'C'}\,{\dot v}^{BB'} v^{CC'}=0.
\end{eqnarray}
This relation can then be written
\begin{eqnarray}
{\dot v}^{BB'} v^{CC'}- {\dot v}^{CB'} v^{BC'}- {\dot v}^{BC'}
v^{CB'} + {\dot v}^{CC'} v^{BB'}=0.\label{eq:28}
\end{eqnarray}
Because $\det(v^{AA'})\neq0$ we know that $v^{AA'}$ has an inverse
$u_{AA'}$ satisfying
\begin{eqnarray}
v^{AA'}u_{BA'}=\delta^A_{\ B}\quad {\rm and}\quad v^{AA'} u_{AB'}
=\delta^{A'}_{\ B'}.
\end{eqnarray}
Therefore, contracting (\ref{eq:28}) with $u_{BB'}$ we obtain
\begin{eqnarray}
(u_{BB'}{\dot v}^{BB'})v^{CC'} + (r-2){\dot v}^{CC'} = 0.
\end{eqnarray}
This equation shows that if ${\dot v}^{CC'}$ were not zero, then
it would have to be proportional to $v^{CC'}$. However, if that
were so, then
\begin{eqnarray}
\varepsilon_{ABC}\,\varepsilon_{A'B'C'}\,{\dot v}^{BB'} v^{CC'}=0
\end{eqnarray}
would imply $\det(v^{CC'})=0$, contrary to the assumption that
$v^{CC'}$ is time-like. It follows that ${\dot v}^{\rm a}=0$. A
similar argument shows that for all $r\geq2$ the geodesic equation
(\ref{eq:26}) implies ${\dot v}^{\rm a}=0$. Hence we have deduced
the following result. Let $y^{\rm a}$ and $z^{\rm a}$ be quantum
space-time points with the property that $y^{\rm a}-z^{\rm a}$ is
time-like and future-pointing. Then the affinely parametrised
geodesic $\gamma$ connecting these points in ${\mathcal H}^{r^2}$
is given by
\begin{eqnarray}
X^{\rm a}(s) = z^{\rm a} + \frac{y^{\rm a}-z^{\rm a}}{[\Delta
(y,z)]^{1/r}}\, s
\end{eqnarray}
for $s\in(-\infty,\infty)$, where
\begin{eqnarray}
\Delta(y,z)=g_{\rm ab\cdots c}(y^{\rm a}-z^{\rm a})(y^{\rm
b}-z^{\rm b})\cdots (y^{\rm c}-z^{\rm c}).
\end{eqnarray}

\section{Conserved quantities}

It is a straightforward exercise to verify that the chronometric
form $\Delta$ for the separation between two points is invariant
when the points of ${\mathcal H}^{r^2}$ are subjected to
transformations of the following type:
\begin{eqnarray}
x^{\,AA'} \to \lambda^{A}_{B} {\bar\lambda}^{A'}_{B'} x^{BB'} +
\beta^{AA'} . \label{eq:33}
\end{eqnarray}
Here $\beta^{AA'}$ represents an arbitrary translation in quantum
space-time, $\lambda^{A}_{B}$ is an element of $SL(r,{\mathbb
C})$, and ${\bar\lambda}^{A'}_{B'}$ is the complex conjugate of
$\lambda^A_B$. The relation of this group of transformations to
the Poincar\'e group in the case $r=2$ should be apparent. Indeed,
one of the attractive features of the extension of space-time
geometry that we are putting forward here is that the
hyper-Poincar\'e group allows such a description, which implies a
wide range of intuitively plausible physical features.

More generally, we remark that the proper hyper-Poincar\'e group
preserves the signature of any space-time interval, whether or not
the interval is degenerate, and hence leaves the causal relations
between events unchanged. We refer to a transformation of the form
\begin{eqnarray}
r^{\rm a}\to L^{\rm a}_{\rm b}r^{\rm b}
\end{eqnarray}
as a `hyper-Lorentz transformation' if
\begin{eqnarray}
L^{\rm a}_{\rm b}= \lambda^{A}_{B} {\bar\lambda}^{A'}_{B'}
\end{eqnarray}
for some element $\lambda^{A}_{B} \in SL(r,{\mathbb C})$. In the
case of Minkowski space ($r=2$) it is well known that a
two-component spinor $\xi^{\bf A}$ can be represented by a complex
null bivector
\begin{eqnarray}
\Lambda^{\rm ab}=\xi^{\bf A}\xi^{\bf B}\varepsilon^{{\bf A}^\prime
{\bf B}^\prime}.
\end{eqnarray}
Conversely, the bivector $\Lambda^{\rm ab}$ determines $\xi^{\bf
A}$ up to the transformation
\begin{eqnarray}
\xi^{\bf A} \to -\xi^{\bf A}.
\end{eqnarray}
This geometrical ambiguity is often referred to as the fundamental
`two-valuedness' of two-component spinors in relativity theory. In
the case of a general quantum space-time of dimension $r^2$, a
hyperspinor $\xi^A$ can be represented by a complex null
$r$-vector (antisymmetric tensor of rank $r$) of the form
\begin{eqnarray}
\Lambda^{\rm ab\cdots c}=\xi^A\xi^B\cdots\xi^C
\varepsilon^{A^\prime B^\prime \cdots C^\prime}.
\end{eqnarray}
The $r$-vector $\Lambda^{\rm ab\cdots c}$ is `null' in the sense
that
\begin{eqnarray}
g_{\rm ab\cdots c}\Lambda^{\rm ap\cdots q}\Lambda^{\rm br\cdots
s}=0.
\end{eqnarray}
Then $\Lambda^{\rm ab\cdots c}$ determines $\xi^A$ up to
transformations of the form
\begin{eqnarray}
\xi^A \to \re^{2\pi \ri k/r}\xi^A,
\end{eqnarray}
where $k=1,2,\ldots,r-1$. Hence we can say that hyperspinors have
a fundamental `$r$-valuedness' (cf. Holm~\cite{Holm4}).

The real dimension of the hyper-Lorentz group is $2r^2-2$, and
thus the real dimension of the hyper-Poincar\'e group is $3r^2-2$.
We observe that the dimension of the hyper-Poincar\'e group grows
linearly with the dimension of the quantum space-time itself,
which is given by $r^2$. This can be contrasted with the dimension
of the group arising if we endow ${\mathbb R}^{r^2}$ with a
standard Lorentzian metric with signature $(1,r^2-1)$. In that
case the associated pseudo-orthogonal group has real dimension
$\frac{1}{2} r^2(r^2-1)$, which together with the translation
group gives a total dimension of $\frac{1}{2}r^2(r^2+1)$. The
parsimonious dimensionality of the hyper-Poincar\'e group arises
from the fact that it preserves the system of causal relations
holding between pairs of points in quantum space-time.

In a flat four-dimensional space-time the symmetries of the
Poincar\'e group are associated with a ten-parameter family of
Killing vectors. Thus, for $r=2$ we have the Minkowski metric
(\ref{eq:7}), and the Poincar\'e group is generated by the
ten-parameter family of vector fields $\xi^{\rm a}$ on ${\mathfrak
M}^4$ satisfying
\begin{eqnarray}
{\mathcal L}_\xi g_{\rm ab}=0,
\end{eqnarray}
where ${\mathcal L}_\xi$ denotes the Lie derivative with respect
to $\xi^{\rm a}$. For any vector field $\xi^{\rm a}$ and any
symmetric tensor field $g_{\rm ab}$ we have
\begin{eqnarray}
{\mathcal L}_\xi g_{\rm ab} = \xi^{\rm c} \nabla_{\!\rm c}g_{\rm
ab} + 2g_{{\rm c}({\rm a}} \nabla_{\!{\rm b})} \xi^{\rm c} .
\end{eqnarray}
If $g_{\rm ab}$ is the metric and $\nabla_{\!\rm a}$ denotes the
associated covariant derivative satisfying
\begin{eqnarray}
\nabla_{\!\rm a}g_{\rm bc}=0,
\end{eqnarray}
we obtain the Killing equation
\begin{eqnarray}
\nabla_{\!({\rm a}}\xi_{{\rm b})}=0,
\end{eqnarray}
where $\xi_{\rm a}=g_{\rm ab}\xi^{\rm b}$. The condition
${\mathcal L}_\xi g_{\rm ab}=0$ therefore implies that $\xi^{\rm
a}$ is a Killing vector.

For $r>2$ we have no Riemannian metric, and the usual relations
between symmetries and Killing vectors are lost. What survives,
however, is of interest. Specifically, to generate a symmetry of
the quantum space-time the vector field $\xi^{\rm a}$ has to
satisfy
\begin{eqnarray}
{\mathcal L}_\xi g_{\rm ab\cdots c} = 0,
\end{eqnarray}
where $g_{\rm ab\cdots c}$ is the chronometric tensor. For a
general vector field $\xi^{\rm a}$ and a general symmetric tensor
field $g_{\rm ab\cdots c}$ we have
\begin{eqnarray}
{\mathcal L}_\xi g_{\rm ab\cdots c} = \xi^{\rm d} \nabla_{\!\rm d}
g_{\rm ab\cdots c} + r\,g_{{\rm d}({\rm a\cdots b}} \nabla_{\!{\rm
c})} \xi^{\rm d} .
\end{eqnarray}
In the case of the quantum space-time ${\mathcal H}^{r^2}$ we let
$\nabla_{\!{\rm a}}$ be the natural flat connection for which
\begin{eqnarray}
\nabla_{\!{\rm a}} g_{\rm bc\cdots d}=0.
\end{eqnarray}
Then to generate a symmetry of the chronometric structure of
${\mathcal H}^{r^2}$ the vector field $\xi^{\rm a}$ has to satisfy
\begin{eqnarray}
g_{{\rm d}({\rm a\cdots b}} \nabla_{\!{\rm c})} \xi^{\rm d} = 0.
\label{eq:4.5}
\end{eqnarray}
Equation (\ref{eq:4.5}) can be written in an alternative form if
we define a symmetric tensor $K_{\rm ab\cdots c}$ of rank $r-1$ by
setting
\begin{eqnarray}
K_{\rm ab\cdots c} = g_{\rm ab\cdots cd}\xi^{\rm d} .
\label{eq:4.6}
\end{eqnarray}
Then (\ref{eq:4.5}) says that $K_{\rm ab\cdots c}$ satisfies the
conditions for a symmetric Killing tensor:
\begin{eqnarray}
\nabla_{\!({\rm a}}K_{\rm bc\cdots d)} = 0 . \label{eq:4.7}
\end{eqnarray}
Thus we see that ${\mathcal H}^{r^2}$ provides an example of a
symmetry group generated by a family of Killing tensors. The
symmetries of the quantum space-time are generated, more
specifically, by a system of $3r^2-2$ irreducible symmetric
Killing tensors of rank $r-1$. The significance of Killing tensors
is that they are associated with conserved quantities. For other
examples of Killing tensors arising in a physical context, see,
e.g., \cite{Hughston8,Hughston9, Hughston10,Penrose9,Walker}. In
the present setting it follows that if the vector field $v^{\rm
a}$ satisfies the geodesic equation, which on a quantum space-time
of dimension $r^2$ is given, as we have seen, by
\begin{eqnarray}
g_{\rm abc\cdots d}\left( v^{\rm e}\nabla_{\!\rm e}v^{\rm
b}\right) v^{\rm c}\cdots v^{\rm d} = 0,
\end{eqnarray}
and if $K_{\rm ab\cdots c}$ is the Killing tensor of rank $r-1$
given by (\ref{eq:4.6}), then we have the following conservation
law:
\begin{eqnarray}
v^{\rm e}\nabla_{\!\rm e}\left( K_{\rm ab\cdots c} v^{\rm a}
v^{\rm b}\cdots v^{\rm c}\right) = 0.
\end{eqnarray}
In other words, the quantity
\begin{eqnarray}
K = K_{\rm ab\cdots c} v^{\rm a}v^{\rm b}\cdots v^{\rm c}
\end{eqnarray}
is a constant of the motion.

\section{Hyper-relativistic mechanics}

It follows from the material of the previous section that in
higher-dimensional quantum space-times the main conservation laws
and symmetry principles of relativistic physics remain intact. In
particular, the conservation of hyper-relativistic momentum and
angular momentum for a system of interacting particles can be
given a well-defined formulation, the basic principles of which
are similar to those applicable in the Minkowskian case.

For this purpose it will be useful to introduce the notion of an
`elementary system' in hyper-relativistic mechanics. Such a system
is defined by its hyper-relativistic momentum and angular
momentum. The hyper-relativistic momentum of an elementary system
is given by a momentum covector $P_{\rm a}$. The associated mass
$m$ is given (cf. \cite{Finkelstein1}) by the following natural
expression:
\begin{eqnarray}
m=\left( g^{\rm ab\cdots c}P_{\rm a}P_{\rm b}\cdots P_{\rm c}
\right)^{\frac{1}{r}}.
\end{eqnarray}
The hyper-relativistic angular momentum of an elementary system is
given by a tensor $L^{\rm b}_{\rm a}$ of the form
\begin{eqnarray}
L^{\rm b}_{\rm a} = l^{B}_{A}\delta^{B'}_{A'} + {\bar l}^{B'}_{A'}
\delta^{B}_{A} ,
\end{eqnarray}
where the hyperspinor $l^{B}_{A}$ is required to be trace-free:
$l^{A}_{A}=0$. The angular momentum is defined with respect to a
choice of origin in such a manner that under a change of origin
defined by a shift vector $\beta^{\rm a}$ we have
\begin{eqnarray}
l^{B}_{A}\to l^B_A + P_{AC'}\beta^{BC'}.
\end{eqnarray}
In the case $r=2$ these formulae reduce to the usual expressions
for relativistic momentum and angular momentum in a Minkowskian
setting. The real covector
\begin{eqnarray}
S_{AA'} = \ri m^{-1} \left( l^B_A P_{A'B} - {\bar l}^{B'}_{A'}
P_{AB'} \right)
\end{eqnarray}
is invariant under a change of origin, and carries the
interpretation of the intrinsic spin of the elementary system. The
magnitude of the spin is then defined by
\begin{eqnarray}
S=|g^{\rm ab\cdots c} S_{\rm a}S_{\rm b}\cdots S_{\rm
c}|^{\frac{1}{r}}.
\end{eqnarray}
In the case of a set of interacting hyper-relativistic systems we
require that the total momentum and angular momentum should be
conserved. This then implies conservation of the total mass and
spin. In short, we see that the idea of `relativistic mechanics'
carries through nicely to the case of a general quantum
space-time.

Now what is the interpretation of these conservation laws? We
shall show later, once we introduce the idea of symmetry breaking,
that hypermomentum can be interpreted as the momentum operator for
a relativistic quantum system. Conservation of hypermomentum then
can be thought of as conservation of four-momentum in relativistic
quantum mechanics in the Heisenberg representation.

\section{Complex null directions}

\begin{figure}
  \includegraphics[height=.4\textheight,angle=270]{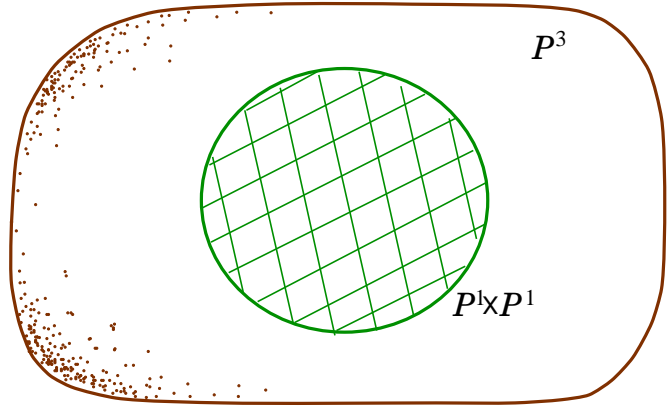}
  \caption{\label{fig:2}{\sl Complex null directions in a
  four-dimensional space-time}. The space of directions at a point
  in complex Minkowski space has the geometry of a complex
  projective 3-space. The complex null directions
  constitute a quadric which has the structure of
  a doubly ruled surface ${\mathbb P}^1\times{\mathbb P}^1$.}
\end{figure}

In four-dimensional space-time it is useful in many contexts to
examine the geometry of the space of complex null vectors at a
point in the space-time. This has the effect of giving us a vivid
picture of the local causal relationships in space-time, and by
sticking with a complex picture we also retain the link with
quantum mechanics. Thus we consider complex vectors $z^{\rm a}$
satisfying the quadratic equation
\begin{eqnarray}
g_{\rm ab}z^{\rm a}z^{\rm b}=0.
\end{eqnarray}
In spinor terms this relation implies that the corresponding
complex matrix $z^{{\bf AA}'}$ is of the special form
\begin{eqnarray}
z^{{\bf AA}'} = \alpha^{\bf A}\beta^{{\bf A}'}. \label{eq:7.1}
\end{eqnarray}
The space of complex vectors at a point in Minkowski space is
${\mathbb C}^4$. The space of complex directions (which results if
we consider equivalence class of vectors modulo overall
proportionality) is the complex projective 3-space ${\mathbb
P}^3$. The null directions constitute a quadric ${\mathbb Q}^2$ in
that space, which owing to the decomposition (\ref{eq:7.1}) has
the structure of a doubly ruled surface
\begin{eqnarray}
{\mathbb Q}^2={\mathbb P}^1 \times {\mathbb P}^1.
\end{eqnarray}
We can identify the first set of lines (the $\alpha$-lines) with
the projective unprimed spinors, and the second set of lines (the
$\beta$-lines) with the projective primed spinors. The quadric
${\mathbb Q}^2$ is ruled in such a way that two lines of the same
type do not intersect, whereas two lines of the opposite type
intersect at a point in ${\mathbb Q}^2$ (see Figure~\ref{fig:2}).
This point corresponds to the null direction they jointly
determine.

In the case of a general quantum space-time of dimension $r^2$, we
consider the space of complex vectors at a point, and examine the
corresponding space of directions, which has the structure of a
complex projective space ${\mathbb P}^{r^2-1}$. The space of
degenerate complex directions, which is given by the vanishing of
the chronometric form $g_{\rm ab \cdots c}z^{\rm a}z^{\rm b}\cdots
z^{\rm c}$, is a hypersurface in ${\mathbb P}^{r^2-1}$, which we
shall call ${\mathcal D}^{r^2-2}$.

The points of ${\mathcal D}^{r^2-2}$ correspond to degenerate
directions of degree $r-1$. The null directions in ${\mathcal
D}^{r^2-2}$ correspond to those directions for which the
associated degenerate vectors are of minimal rank and hence of the
form $z^{AA'}=\alpha^A\beta^{A'}$. These constitute a subvariety
${\mathbb Q}^{2r-2}\subset{\mathcal D}^{r^2-2}$ defined by the
mutual intersection of a system of quadrics, given by the equation
\begin{eqnarray}
g_{\rm ab \cdots c}z^{\rm a} z^{\rm b}=0.
\end{eqnarray}
In this situation we have
\begin{eqnarray}
{\mathbb Q}^{2r-2}={\mathbb P}^{r-1} \times {\mathbb P}^{r-1},
\end{eqnarray}
and we can identify the two systems of $(r-1)$-planes by which
${\mathbb Q}^{2r-2}$ is foliated, which we refer to as
$\alpha$-planes and $\beta$-planes, as the spaces of projective
unprimed and primed hyperspinors, respectively. The various
degenerate directions of intermediate degree correspond to points
in ${\mathcal D}^{r^2-2}$ lying on the linear spaces spanned by
the joins of $d$ points in ${\mathbb Q}^{2r-2}$
($d=2,3,\ldots,r$). The degree of degeneracy is given by the
integer $d$.

\begin{figure}
  \includegraphics[height=.4\textheight,angle=270]{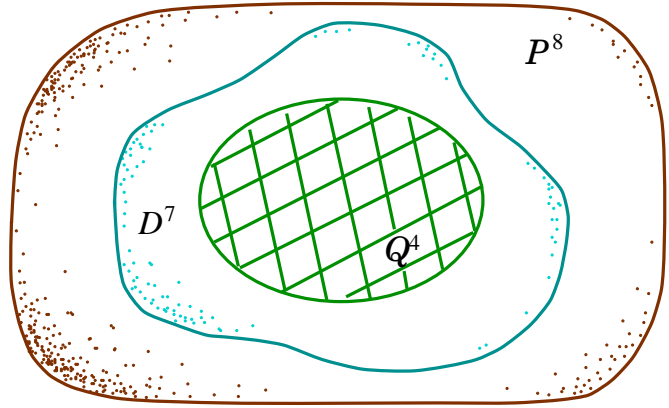}
  \caption{\label{fig:3}{\sl The space of complex null directions
  at a point in a quantum space-time of dimension nine}. In this
  case the space of degenerate directions is given by a cubic
  surface ${\mathcal D}^{7}$ in the complex projective space
  ${\mathbb P}^8$. The complex null directions then constitute a
  four-dimensional doubly foliated subvariety ${\mathbb Q}^4=
  {\mathbb P}^2\times {\mathbb P}^2$ in ${\mathcal D}^{7}$.}
\end{figure}

In the case $r=3$, illustrated in Figure~\ref{fig:3}, the space of
complex directions at a point in $\mathcal{C\!H}^9$ is ${\mathbb
P}^{8}$, and the degenerate directions constitute a cubic
hypersurface ${\mathcal D}^{7}\subset {\mathbb P}^8$. The null
directions lie in the doubly foliated surface
\begin{eqnarray}
{\mathbb Q}^4 ={\mathbb P}^{2} \times {\mathbb P}^{2}
\end{eqnarray}
in ${\mathcal D}^{7}$. The points of ${\mathcal D}^{7}$ all lie on
the `first join' of ${\mathbb Q}^4$ with itself; in other words,
any point of ${\mathcal D}^7$ lies on a line joining two points of
${\mathbb Q}^4$. Thus we write ${\mathcal D}^7=J_1({\mathbb
Q}^4)$. The space ${\mathcal D}^7\!\setminus\!{\mathbb Q}^4$ then
consists of degenerate directions that are strictly of the second
degree. Note that any point of ${\mathbb P}^8$ can be represented
as the join of three points in ${\mathbb Q}^4$, and hence $J_2(
{\mathbb Q}^4)={\mathbb P}^8$.

\begin{figure}
  \includegraphics[height=.5\textheight,angle=270]{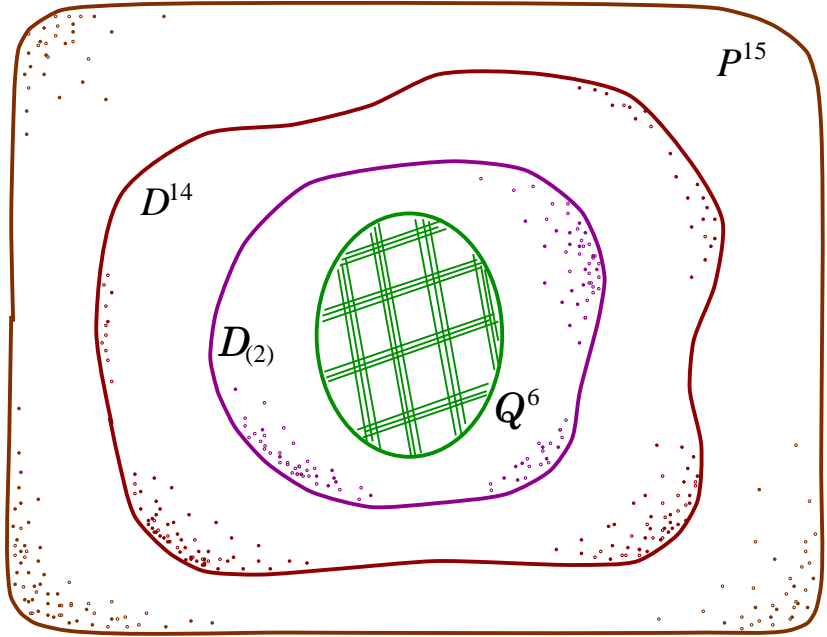}
  \caption{\label{fig:4}{\sl The space of complex null directions
  at a point in
  the quantum space-time} ${\mathcal H}^{16}$. The degenerate
  directions of degree three constitute a quartic hypersurface
  ${\mathcal D}^{14}$ in ${\mathbb P}^{15}$.
  The null directions belong to a six-dimensional doubly foliated
  subvariety ${\mathbb Q}^6={\mathbb P}^3\times{\mathbb P}^3$.
  Degenerate directions of degree two constitute a submanifold
  ${\mathcal D}_{(2)}=J_1({\mathbb Q}^{6})\subset{\mathcal D}^{14}$
  consisting of points that lie on lines spanned by pairs of
  points in the subvariety ${\mathbb Q}^6$.}
\end{figure}

In the case $r=4$, illustrated in Figure~\ref{fig:4}, the space of
complex directions at a point in $\mathcal{C\!H}^{16}$ is the
complex projective space ${\mathbb P}^{15}$, and the degenerate
directions constitute a quartic hypersurface ${\mathcal D}^{14}
\subset {\mathbb P}^{15}$. The null directions (degenerate
direction of the first degree) lie on the doubly foliated surface
\begin{eqnarray}
{\mathcal D}_{(1)}={\mathbb P}^{3} \times {\mathbb P}^{3}
\end{eqnarray}
in ${\mathcal D}^{14}$. The degenerate directions of the second
degree lie on the first join of ${\mathbb Q}^6$ with itself:
\begin{eqnarray}
{\mathcal D}_{(2)}=J_1({\mathbb Q}^{6}).
\end{eqnarray}
The degenerate directions of the third degree lie in
\begin{eqnarray}
{\mathcal D}_{(3)} =J_2({\mathbb Q}^{6})
\end{eqnarray}
and constitute the general elements of ${\mathcal D}^{14}$.

\section{Twistors and hypertwistors}

The proceeding discussion of complex null directions can be seen
as a `warm-up' exercise for the introduction of the concept of
hypertwistors. For this purpose we introduce a notation that
closely parallels the standard notation for $r=2$ Penrose
twistors. Let us denote by ${\mathbb T}^{\boldsymbol\alpha}$ the
complex vector space of dimension $2r$ given by the pair
$({\mathbb S}^{A},{\mathbb S}_{A'})$. We write
\begin{eqnarray}
Z^{\alpha}= (\omega^{A},\pi_{A'})
\end{eqnarray}
for a typical element of ${\mathbb T}^{{\alpha}}$. Such elements
will be referred to as `hypertwistors' (also called `generalised
twistors' \cite{Eastwood,Hughston1}). Let ${\mathbb
T}_{{\alpha}}=({\mathbb S}_{A},{\mathbb S}^{A'})$ denote the space
of dual hypertwistors. A natural pseudo-Hermitian structure can be
introduced on the geometry of hypertwistors by means of the
complex conjugation operation that maps $(\omega^{A},\pi_{A'})
\in{\mathbb T}^{{\alpha}}$ to $({\bar\pi}_{A},{\bar\omega}^{A'})
\in{\mathbb T}_{{\alpha}}$. The corresponding pseudo-Hermitian
form is then given by
\begin{eqnarray}
Z^{{\alpha}} {\bar Z}_{{\alpha}} = \omega^{A} {\bar\pi}_{A} +
\pi_{A'} {\bar\omega}^{A'}.
\end{eqnarray}
It is straightforward to verify that the inner product
$Z^{\alpha}{\bar Z}_{{\alpha}}$ is invariant under the action of
the group $U(r,r)$. In the case $r=2$, the elements of ${\mathbb
T}^{\boldsymbol\alpha}$ are standard Penrose twistors, and the
relevant symmetry group is $U(2,2)$.

\begin{figure}
  \includegraphics[height=.4\textheight,angle=270]{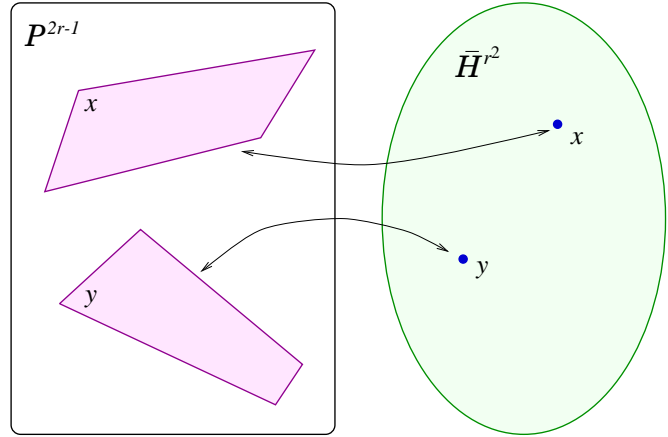}
  \caption{\label{fig:5}{\sl The Klein representation for a
  general quantum space-time}. The aggregate consisting of all
  $(r-1)$-planes
  in the complex projective space ${\mathbb P}^{2r-1}$ forms a
  manifold of dimension $r^2$. This space is the complexified,
  compactified version of the quantum space-time ${\mathcal
  C}\!{\mathcal H}^{r^2}$.}
\end{figure}

The space ${\mathbb P}^{2r-1}$ of projective hypertwistors is a
good starting point for the analysis of the conformal geometry of
complex quantum space-time, which can be regarded as the
Grassmannian variety of projective $(r-1)$-planes in ${\mathbb
P}^{2r-1}$, as illustrated in Figure~\ref{fig:5}. More precisely,
the aggregate of all projective $(r-1)$-planes in ${\mathbb
P}^{2r-1}$ constitutes a compact manifold of dimension $r^2$,
which we identify as the complexified, compactified quantum
space-time ${\mathcal C}\!\bar{\mathcal H}^{r^2}$. The `finite'
points of ${\mathcal C}\!\bar{\mathcal H}^{r^2}$ correspond to
those $(r-1)$-planes of ${\mathbb P}^{2r-1}$ that are determined
by a linear relation of the form
\begin{eqnarray}
\omega^{A} = \ri x^{\,AA'} \pi_{A'} \label{eq:54}
\end{eqnarray}
for some fixed $x^{\,AA'}$. Thus for each (complex) point
$x^{\,AA'}$ in $\mathcal{C\!H}^{r^2}$ we obtain, according to
equation (\ref{eq:54}), an $(r-1)$-plane in ${\mathbb P}^{2r-1}$.
The aggregate of such $(r-1)$-planes constitute the points of
$\mathcal{C\!H}^{r^2}$. The $(r-1)$-planes for which $x^{\,AA'}$
is Hermitian then constitute the points of the real quantum
space-time ${\mathcal H}^{r^2}$.

\begin{figure}
  \includegraphics[height=.5\textheight,angle=270]{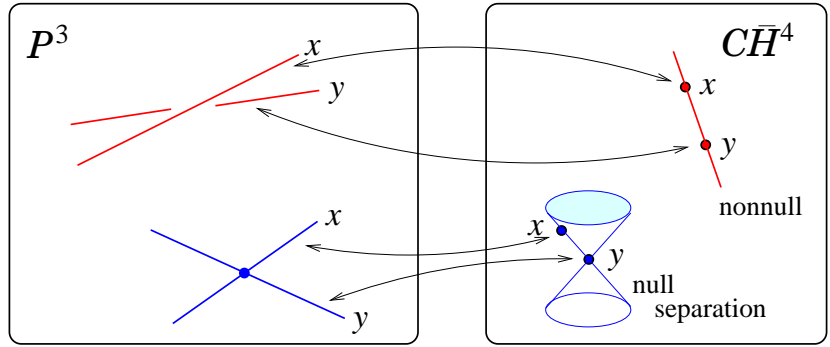}
  \caption{\label{fig:6}{\sl Twistor theory}. The lines of the
  complex projective space ${\mathbb P}^3$ correspond to the
  points of the complex space-time ${\mathcal C}\!{\bar{\mathcal
  H}}^4$. Two points in ${\mathcal C}\!\bar{\mathcal H}^4$ are
  null-separated if and only if the corresponding lines in
  ${\mathbb P}^3$ intersect.}
\end{figure}

\begin{figure}
  \includegraphics[height=.6\textheight,angle=270]{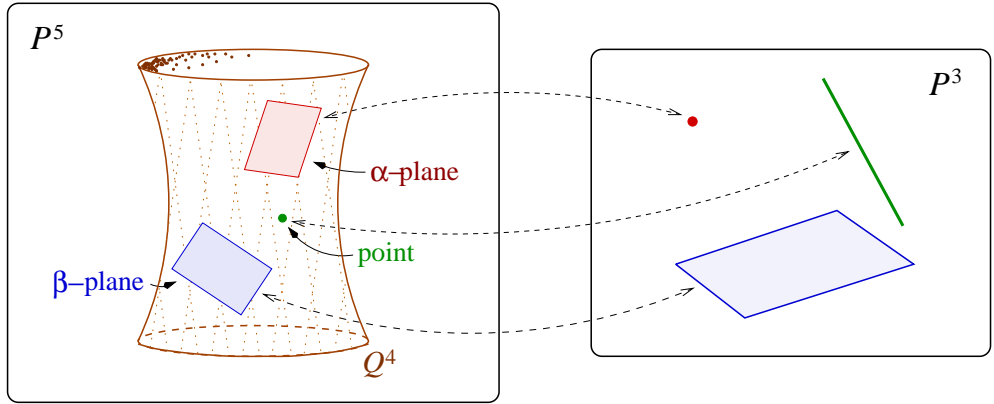}
  \caption{\label{fig:7}{\sl Complexified, compactified Minkowski
  space-time as the Klein quadric in complex projective
  5-space.} The aggregate of complex projective lines in
  ${\mathbb P}^3$ constitutes a nondegenerate quadric $Q^4$ in
  ${\mathbb P}^5$. The quadric contains two distinct systems of
  projective 2-planes, called $\alpha$-planes and
  $\beta$-planes. Any two distinct planes of the same type in
  $Q^4$ intersect at a point. Two planes of the opposite type in
  $Q^4$ will in general not intersect, but if they do, they
  intersect in a line. The points of ${\mathbb P}^3$ correspond to
  $\alpha$-planes in $Q^4$, and the 2-planes of ${\mathbb P}^3$
  correspond to $\beta$-planes in $Q^4$.}
\end{figure}

The conformal structure of a quantum space-time is implicit in the
various possibilities arising for the intersections of
$(r-1)$-planes in the projective hypertwistor space ${\mathbb
P}^{2r-1}$. In the case $r=2$ (Penrose twistors) the projective
space ${\mathbb P}^3$ contains a four-dimensional family of
complex projective lines, the aggregation of which constitutes the
associated space-time (see Figure~\ref{fig:6}). In this case, a
pair of space-time points $x$ and $y$ are null-separated in the
Minkowskian sense if and only if the corresponding lines in
${\mathbb P}^3$ intersect. The space of all complex projective
lines in ${\mathbb P}^3$ constitutes a four-dimensional quadric
hypersurface $Q^4$ in ${\mathbb P}^5$, which we identify as
complexified, compactified Minkowski space-time ${\mathcal
C}\!\bar{\mathcal H}^{4}$. The quadric $Q^4$ contains two systems
of projective 2-planes, called $\alpha$-planes and $\beta$-planes.
The $\alpha$-planes are in one-to-one correspondence with the
points of ${\mathbb P}^3$, whereas the $\beta$-planes are in
one-to-one correspondence with the 2-planes of ${\mathbb P}^3$, as
illustrated in Figure~\ref{fig:7}.

\begin{figure}
  \includegraphics[height=.6\textheight,angle=270]{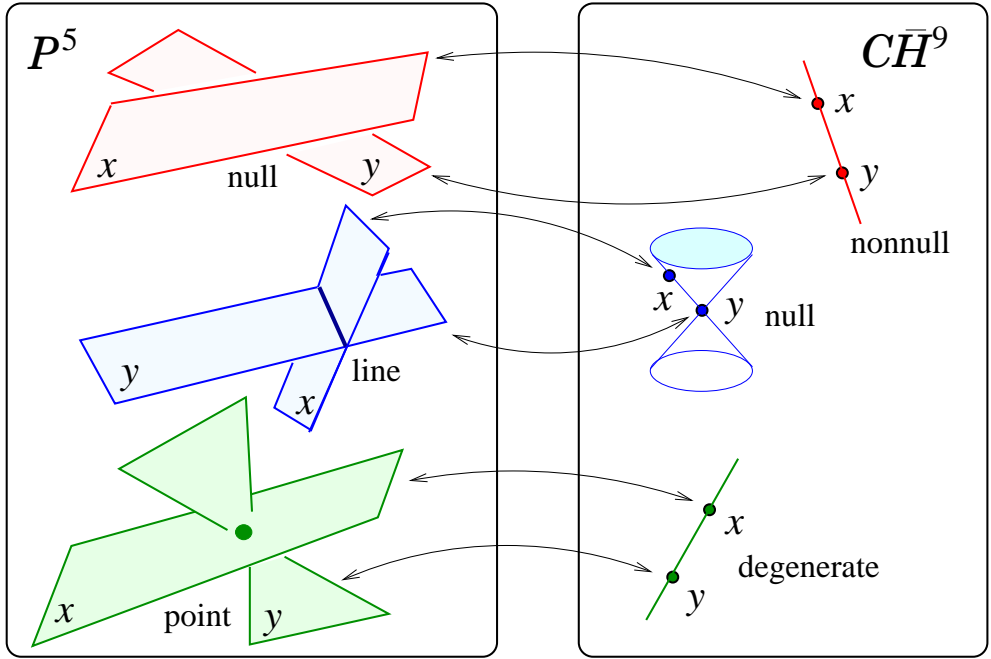}
  \caption{\label{fig:8}{\sl Hypertwistor theory for $r=3$}. The
  aggregate of complex projective $2$-planes in ${\mathbb P}^5$
  constitutes a nine-dimensional quantum space-time
  ${\mathcal C}\!{\bar{\mathcal H}}^{9}$. In general, two such
  planes will
  not intersect. If two planes intersect in a point, then the
  separation vector of the corresponding space-time points is
  degenerate. If the intersection is in a line, then
  the corresponding points are null-separated.}
\end{figure}

The case $r=3$ has been studied closely by Finkelstein~\cite{
Finkelstein1} and his collaborators. In this case, the
nine-dimensional quantum space-time ${\mathcal C}\!\bar{\mathcal
H}^{9}$ arises when we consider the Grassmanian of complex
projective 2-planes in complex projective 5-space. A pair of
planes in ${\mathbb P}^3$ in general will not intersect. If they
do, however, then there are three possibilities: they can
intersect in a point, a line, or a plane. These situations
correspond to the various levels of degeneracy obtainable for the
separation vector. In particular, if the two planes intersect in a
line, then the separation vector is null (see Figure~\ref{fig:8}).

A pair of $(r-1)$-planes in ${\mathbb P}^{2r-1}$ will not in
general intersect. This generic non-intersection property
corresponds to the nonvanishing of the chronometric form for the
separation vector of generic quantum space-time points. An
$(r-1)$-plane in ${\mathbb P}^{2r-1}$ is represented by a simple
skew hypertwistor $P^{{\alpha}{\beta}\cdots{\gamma}}$ of rank $r$.
By a `simple' skew hypertwistor we mean one of the form
\begin{eqnarray}
P^{{\alpha}{\beta}\cdots{\gamma}} = A^{[{\alpha}}B^{{\beta}}\cdots
C^{{\gamma}]}
\end{eqnarray}
for some set $A^{\alpha},B^{\alpha},\cdots,C^{\alpha}$ of $r$
hypertwistors. Suppose that the simple skew hypertwistors
$P^{{\alpha}{\beta} \cdots{\gamma}}$ and
$Q^{{\alpha}{\beta}\cdots{\gamma}}$ represent, respectively, the
$(r-1)$-planes $P$ and $Q$ in ${\mathbb P}^{2r-1}$. A necessary
and sufficient condition for the vanishing of the chronometric
form for the corresponding space-time points is
\begin{eqnarray}
\varepsilon_{\alpha\beta\cdots\gamma\rho\sigma\cdots\tau}
P^{\alpha\beta\cdots\gamma} Q^{\rho\sigma\cdots\tau} = 0,
\label{eq:5.4}
\end{eqnarray}
where $\varepsilon_{{\alpha}{\beta}\cdots{\gamma}{\rho}{\sigma}
\cdots{\tau}}$ is the skew hypertwistor of rank $2r$. We note that
(\ref{eq:5.4}) is symmetric under the interchange of $P$ and $Q$
if $r$ is even, and antisymmetric if $r$ is odd. The vanishing of
(\ref{eq:5.4}) is the condition that the planes $P$ and $Q$
contain a point in common. This means that the skew hypertwistors
$P^{{\alpha}{\beta}\cdots{\gamma}}$ and $Q^{{\rho}{\sigma}
\cdots{\tau}}$ contain at least one hypertwistor as a common
factor. Thus, a necessary and sufficient condition for a pair of
quantum space-time points to have a degenerate separation is that
the corresponding $(r-1)$-planes in ${\mathbb P}^{2r-1}$ should
intersect.

More generally, the degeneracy $d$ of the separation of a pair of
quantum space-time events is given by $d=r-m-1$, where $m$ is the
dimensionality of the intersection of the corresponding
$(r-1)$-planes in ${\mathbb P}^{2r-1}$. The possible degrees of
degeneracy are $d=1,2,\ldots,r-1$. If we interpret the case of no
intersection as an intersection of dimension $-1$, then a
nondegenerate separation can be interpreted as a `degeneracy of
degree $r$'. Thus separations of degree less than $r$ are
degenerate, whereas a separation of degree $r$ is nondegenerate.
The degree of degeneracy is given by the rank of the separation
matrix $r^{\,AA'}=x^{\,AA'}-y^{\,AA'}$.

Alternatively, given two skew hypertwistors $P^{{\alpha}{\beta}
\cdots{\gamma}}$ and $Q^{{\alpha}{\beta} \cdots{\gamma}}$, each
with $r$ indices, let us form the dual hypertwistor by
\begin{eqnarray}
Q_{{\alpha}{\beta}\cdots{\gamma}} =\varepsilon_{{\alpha}{\beta}
\cdots {\gamma}{\rho}{\sigma} \cdots{\tau}}Q^{{\rho}
{\sigma}\cdots{\tau}}.
\end{eqnarray}
Then $d$ is the maximum number of index contractions we can make
between $P^{{\alpha}{\beta} \cdots{\gamma}}$ and
$Q_{{\alpha}{\beta}\cdots{\gamma}}$ without obtaining the result
zero. If a single index contraction gives zero, this corresponds
to the case where $P^{{\alpha}{\beta} \cdots{\gamma}}$ is
proportional to $Q^{{\alpha}{\beta} \cdots{\gamma}}$. Thus $d=0$
(separation of degree zero) can be interpreted as the `completely
degenerate' case where the two space-time points coincide.

\section{Hypertwistor fields}

Now we present a few elementary applications of hypertwistor
theory. In the case of a four-dimensional space-time, it is well
known that twistors can be characterised as solutions of the
differential equation
\begin{eqnarray}
\nabla^{{\bf A}'({\bf A}}\xi^{{\bf B})}=0, \label{eq:73}
\end{eqnarray}
the so-called `twistor equation'. Indeed, the solution of this
equation is
\begin{eqnarray}
\xi^{\bf A}=\omega^{\bf A}-{\rm i} x^{{\bf AA}'} \pi_{{\bf A}'},
\end{eqnarray}
and the associated twistor is determined by the fixed spinor pair
$(\omega^{\bf A}, \pi_{{\bf A}'})$. More generally, in the case of
a quantum space-time of dimension $r^2$ we consider the equation
\begin{eqnarray}
r \nabla_{AA'}\xi^C = \delta^C_A \nabla_{DA'}\xi^D .
\label{eq:2.1}
\end{eqnarray}
It is not difficult to verify that in the case $r=2$ equation
(\ref{eq:2.1}) reduces to the conventional twistor equation
(\ref{eq:73}). The solutions of (\ref{eq:2.1}) can be expressed in
terms of generalised twistors:
\begin{eqnarray}
\xi^A = \omega^A - {\rm i}x^{\,AA'}\pi_{A'} , \label{eq:2.2}
\end{eqnarray}
where $\omega^A$ and $\pi_{A'}$ are constant. This result is
established in \cite{Hughston1}.

We note that a necessary and sufficient condition for a
hypertwistor field $\xi^A(x)$ to satisfy the hypertwistor equation
(\ref{eq:2.1}) is that it should satisfy
\begin{eqnarray}
\sigma_{C}\sigma_{[A}\nabla_{B]C'}\xi^C=0
\end{eqnarray}
for all values of $\sigma_A$. For $r=2$ the twistor equation
(\ref{eq:73}) is a special case of a more general nonlinear
partial differential equation known as the geodesic shear-free
condition:
\begin{eqnarray}
\xi^{\bf A}\xi^{\bf B}\nabla_{{\bf A}'{\bf A}}\xi_{\bf B}=0.
\label{eq:78}
\end{eqnarray}
When $r>2$, however, the connection between the twistor equation
and the geodesic shear-free condition is lost, and the appropriate
generalisation of the latter is
\begin{eqnarray}
\left( \xi_{[A}\nabla_{B]B'}\xi_{[C}\right) \xi_{D]}=0,
\label{eq:79}
\end{eqnarray}
which reduces to the geodesic shear-free condition (\ref{eq:78})
when $r=2$. Solutions of the equation (\ref{eq:79}) can be
generated through the consideration of the intersections of
projective $(r-1)$-planes in ${\mathbb P}^{2r-1}$ with certain
types of complex analytic varieties in ${\mathbb P}^{2r-1}$.

Now let us consider the dynamical equations for fields of totally
null momentum on a general quantum space-time. The relevant field
equations take the form
\begin{eqnarray}
\nabla_{A[A'} \nabla_{B']B} \phi = 0, \label{eq:1.43}
\end{eqnarray}
\begin{eqnarray}
\nabla_{A[A'} \phi_{B']\cdots C'} = 0, \label{eq:1.42}
\end{eqnarray}
and
\begin{eqnarray}
\nabla_{A'[A} \phi_{B]\cdots C} = 0. \label{eq:1.425}
\end{eqnarray}
These equations generalise the zero rest-mass equations for each
helicity $s=\pm\frac{1}{2}n$ (where $n=0,1,2,\ldots$) in
four-dimensional space-time.

The solutions can be described briefly as follows. We write
$Z^\alpha =(\omega^A,\pi_{A'})$ and let $\psi(Z^\alpha)$ denote an
analytic, homogeneous function of degree $-r+2s$ defined on some
region of the hypertwistor space. For $s=0$ we set
\begin{eqnarray}
\phi(x) = \oint \psi(\ri x^{\,AA'}\pi_{A'},\pi_{A'})
\vartriangle\!\!\pi , \label{eq:3.2}
\end{eqnarray}
where the differential form $\vartriangle\!\!\pi$ is given by
\begin{eqnarray}
\vartriangle\!\!\pi = \epsilon^{A'B'\cdots C'} \pi_{A'}\pi_{B'}
\cdots \pi_{C'} . \label{eq:3.3}
\end{eqnarray}
Note that $\vartriangle\!\!\pi$ is homogeneous of degree $r$ in
$\pi_{A'}$, and thus that the quantity appearing in the integral
sign in (\ref{eq:3.2}) is homogeneous of degree zero. Similarly,
for helicity $s=-\frac{1}{2}$ we define
\begin{eqnarray}
\phi_{A'}(x) = \oint \pi_{A'}\psi(\ri x^{\,AA'}\pi_{A'}, \pi_{A'})
\vartriangle\!\!\pi , \label{eq:3.25}
\end{eqnarray}
and for helicity $s=\frac{1}{2}$ we define
\begin{eqnarray}
\phi_{A'}(x) = \oint \frac{\partial}{\partial\omega^A} \psi(\ri
x^{\,AA'}\pi_{A'}, \pi_{A'}) \vartriangle\!\!\pi .
\label{eq:3.255}
\end{eqnarray}
Here it is understood that first we differentiate $\psi(
\omega^{A}, \pi_{A'})$ with respect to $\omega^A$, and we set
$\omega^A=\ri x^{\,AA'}\pi_{A'}$ before integrating. A proposition
established in \cite{Hughston1} shows that the contour integrals
(\ref{eq:3.2}), (\ref{eq:3.25}), and (\ref{eq:3.255}) satisfy the
totally null momentum conditions (\ref{eq:1.43}), (\ref{eq:1.42}),
and (\ref{eq:1.425}), respectively.

As an illustration, let us first take the simplest case, given by
equation (\ref{eq:3.2}) for $r=3$. For our hypertwistor function
we take
\begin{eqnarray}
\psi(Z^\alpha) = \frac{1}{P_\alpha Z^\alpha Q_\beta Z^\beta
R_\gamma Z^\gamma}  , \label{eq:4.1}
\end{eqnarray}
where $P_\alpha$, $Q_\beta$, and $R_\gamma$ denote fixed points in
the dual hypertwistor space. For their hyperspinor decomposition
we write
\begin{eqnarray}
P_\alpha = ( p_A,p^{\,A'}), \quad Q_\alpha = ( q_A,q^{\,A'}),
\quad {\rm and} \quad  R_\alpha = ( r_A,r^{\,A'}), \label{eq:4.2}
\end{eqnarray}
and set
\begin{eqnarray}
{\tilde p}^{\,A'}(x) = p^{\,A'} + {\ri}x^{\,A'A} p_A ,\quad
{\tilde q}^{\,A'}(x) = q^{\,A'} + {\ri}x^{\,A'A}q_A,\quad {\rm
and} \quad {\tilde r}^{\,A'}(x) = r^{\,A'} + {\ri}x^{\,A'A}r_A ,
\label{eq:4.3}
\end{eqnarray}
where ${\tilde p}^{\,A'}$, ${\tilde q}^{\,B'}$, and ${\tilde
r}^{\,C'}$ are solutions of the primed analogue
\begin{eqnarray}
r\nabla_{AA'}{\tilde\eta}^{C'} = \delta^{C'}_{A'} \nabla_{AD'}
{\tilde\eta}^{D'} , \label{eq:2.9}
\end{eqnarray}
of the hypertwistor equation (\ref{eq:2.1}). Inserting this
expression into the contour integral formula, we obtain the result
\begin{eqnarray}
\phi(x) = \frac{4\pi^2}{\varepsilon_{A'B'C'} {\tilde
p}^{\,A'}{\tilde q}^{\,B'}{\tilde r}^{\,C'}},  \label{eq:4.66}
\end{eqnarray}
where the contour is taken to be $S^1\times S^1$. It is then a
straightforward exercise to verify directly that (\ref{eq:4.66})
satisfies the scalar hyper-wave equation (\ref{eq:1.43}). This
example generalises the so-called `elementary states' of standard
Penrose twistor theory~\cite{Penrose10}.

\section{Geometrical structures in cosmology}

One of the motivations for the idea of quantum space-time is that
it provides a possible way forward towards the unification of
cosmology and elementary particle physics. Clearly some such
unification is required for a coherent discussion of the early
stages of the universe---but in what mathematical framework should
this unification be pursued? An added impetus to these
considerations comes from the dark matter/energy problem (see,
e.g., \cite{Ellis,Rees} and references cited therein), which has
had the effect of encouraging physicists to rethink the
foundations of cosmology. If we are to consider new classes of
models, then some criteria are required to limit the range of
possibilities. In particular, it is reasonable to propose that:
(a) the model should be geometrical in character; and (b) the
model should be rich enough to admit within its space the
possibility of a geometrisation of quantum mechanics (by
`geometrisation' we mean an approach in the spirit, e.g., of
\cite{Anandan,Ashtekar,Brody1,Gibbons,Kibble}). Our intention now
is to put forward a tentative approach to `quantum cosmology' in
this spirit, based on the idea of quantum space-time.

In doing so, we bear in mind that there are a number of distinct
inter-related geometrical structures arising in cosmology, that
need to be taken into consideration as we proceed. These include:
(a) reality structure (as in the distinction between a real and a
complex space-time); (b) causality structure (identification of
the light cones); (c) infinity structure (is the universe open or
closed?); (d) chronometric structure (how are time, distance, and
energy measured?); and (e) singularity structure (how does one
characterise the beginning and end?).

The theory of hyperspin is well suited for addressing these issues
in a mathematically compelling way (see \cite{Finkelstein3,
Holm2,Holm3} for earlier examples of models for hypercosmologies).
It is convenient to begin the present analysis with a discussion
of the structure of infinity in the case of a general flat quantum
space-time. This will lead us to more general cosmological
considerations.

As indicated above, for any skew hypertwistor $Q^{{\alpha}{\beta}
\cdots{\gamma}}$  of rank $r$ in a quantum space-time of dimension
$r^2$ we define its dual $Q_{{\alpha} {\beta}\cdots{\gamma}}$ by
the relation
\begin{eqnarray}
Q_{{\alpha}{\beta}\cdots{\gamma}} =
\varepsilon_{{\alpha}{\beta}\cdots{\gamma}
{\rho}{\sigma}\cdots{\tau}} Q^{{\rho}{\sigma}\cdots{\tau}} .
\end{eqnarray}
Here $\varepsilon_{{\alpha}{\beta}\cdots{\gamma} {\rho}{\sigma}
\cdots{\tau}}$ is the skew hypertwistor of rank $2r$, which is
unique up to scale. Depending on whether $r$ is even or odd, we
have the following interchange relations:
\begin{eqnarray}
\varepsilon_{{\alpha}{\beta}\cdots{\gamma}{\rho}{\sigma}\cdots
{\tau}} = \pm\varepsilon_{{\rho}{\sigma} \cdots{\tau} {\alpha}
{\beta} \cdots {\gamma}}\, .
\end{eqnarray}
Thus if $r$ is even, then once the scale is fixed we obtain a
symmetric inner product on the space of skew hypertwistors of rank
$r$, which we denote
\begin{eqnarray}
\langle P,Q\rangle = \varepsilon_{{\alpha}
{\beta}\cdots{\gamma}{\rho}{\sigma}\cdots{\tau}}\,
P^{{\alpha}{\beta}\cdots{\gamma}} Q^{{\rho}{\sigma}\cdots{\tau}} .
\label{eq:6.3}
\end{eqnarray}
If $r$ is odd then the product (\ref{eq:6.3}) gives us a
symplectic structure.

Under complex conjugation the skew hypertwistor $P^{{\alpha}
{\beta}\cdots{\gamma}}$ becomes ${\bar P}_{{\alpha} {\beta}\cdots
{\gamma}}$. If $P^{{\alpha}{\beta}\cdots{\gamma}}$ is simple, thus
corresponding to an $(r-1)$-plane $P$ in ${\mathbb P}^{2r-1}$,
then we say that $P$ is real if ${\bar P}_{{\alpha}{\beta} \cdots
{\gamma}}$ is proportional to $P_{{\alpha}{\beta}\cdots
{\gamma}}$. The real $(r-1)$-planes of ${\mathbb P}^{2r-1}$
correspond to the real points of quantum space-time.

The structure at infinity in the compactified quantum space-time
$\bar{\mathcal H}^{r^2}$ can be described as follows. In the
hypertwistor space ${\mathbb P}^{2r-1}$ we choose a real
$(r-1)$-plane $I$ represented by a simple skew hypertwistor
$I^{{\alpha}{\beta}\cdots{\gamma}}$. The point $I$ in
$\bar{\mathcal H}^{r^2}$ corresponding to the $(r-1)$-plane $I$ in
${\mathbb P}^{2r-1}$ will be called the `point at infinity'. The
locus in $\bar{\mathcal H}^{r^2}$ consisting of all points that
have a degenerate separation from $I$ will be called `infinity'.
It should be evident that infinity has a rich structure, with
various domains that can be classified according to the degree of
degeneracy of their separation from the point $I$.

The `finite' points of $\bar{\mathcal H}^{r^2}$ are those for
which the separation from $I$ is nondegenerate, i.e. those points
$P$ for which $\langle P,I\rangle\neq0$. In the case of two finite
quantum space-time points the chronometric form $\Delta$ is given
as follows:
\begin{eqnarray}
\Delta(P,Q) = \frac{
\varepsilon_{{\alpha}{\beta}\cdots{\gamma}{\rho}{\sigma}\cdots{\tau}}
P^{{\alpha}{\beta}\cdots{\gamma}} Q^{{\rho}{\sigma}\cdots{\tau}}}
{(\varepsilon_{{\alpha}{\beta}\cdots{\gamma}{\rho}{\sigma}\cdots{\tau}}
P^{{\alpha}{\beta}\cdots{\gamma}} I^{{\rho}{\sigma}\cdots{\tau}})
(\varepsilon_{{\alpha}{\beta}\cdots{\gamma}{\rho}{\sigma}\cdots{\tau}}
Q^{{\alpha}{\beta}\cdots{\gamma}} I^{{\rho}{\sigma}\cdots{\tau}})}
. \label{eq:6.4}
\end{eqnarray}
Equivalently we can write
\begin{eqnarray}
\Delta(P,Q)=\frac{\langle P,Q\rangle}{\langle P,I\rangle \langle
Q,I\rangle} .
\end{eqnarray}
If $P$ and $I$ are not null-separated, then we can choose the
scales of $P^{{\alpha}{\beta}\cdots{\gamma}}$ and
$I^{{\alpha}{\beta}\cdots{\gamma}}$ such that $\langle P, I\rangle
=1$, without loss of generality, and similarly for
$Q^{{\alpha}{\beta}\cdots{\gamma}}$ and
$I^{{\alpha}{\beta}\cdots{\gamma}}$.

We note that $\Delta(P,Q)$ is independent of the scale of
$P^{{\alpha}{\beta}\cdots{\gamma}}$ and
$Q^{{\alpha}{\beta}\cdots{\gamma}}$. On the other hand,
$\Delta(P,Q)$ does depend on the scale of
$\varepsilon_{{\alpha}{\beta}\cdots{\gamma}{\rho}{\sigma}\cdots{\tau}}$
and the scale of $I^{{\alpha}{\beta}\cdots{\gamma}}$. It has an
epsilon `weight' of $-1$ and an $I$ `weight' of $-2$ (cf.
\cite{Hughston7}). If we form the ratio associated with four
hypertwistors $P$, $Q$, $R$, and $S$, given by
\begin{eqnarray}
\frac{\Delta(P,Q)}{\Delta(R,S)} = \frac{\|p-q\|^r}{\|r-s\|^r},
\label{eq:6.5}
\end{eqnarray}
where $p$, $q$, $r$, and $s$ are the quantum space-time points
corresponding to $P$, $Q$, $R$, and $S$, respectively, then we
obtain an expression that is absolute---that is to say, a
geometric invariant. This is because $\Delta(P,Q)$ has the
`dimensionality' of time raised to the power $r$; whereas the
ratio (\ref{eq:6.5}) arises as a comparison of two such time
intervals, and thus is dimensionless. The basic chronometric
geometry, with infinity chosen as indicated above, admits no
absolute or `preferred' unit of time: in this geometry only ratios
of time intervals have an absolute meaning.

\section{Higher-dimensional twistor cosmology}

There is, on the other hand, no reason {\it a priori} why just
such a structure should apply at infinity. Other choices are
available for $I^{{\alpha}{\beta}\cdots{\gamma}}$, and these have
the effect of giving ${\bar{\mathcal H}}^{r^2}$ the structure of a
cosmological model. In the case $r=2$, for example, if
$I^{\alpha\beta}$ is chosen to be real and non-simple, then the
quadratic form
\begin{eqnarray}
\lambda = \varepsilon_{\alpha\beta \gamma\delta} I^{\alpha\beta}
I^{\gamma\delta}, \label{eq:105}
\end{eqnarray}
which has an epsilon weight of one and an $I$-weight of two, has
the dimensionality of inverse squared-time. Hence in this case
there \emph{is} a preferred unit of time. Other time intervals can
then be expressed in multiples of the preferred unit of time.

\begin{figure}
  \includegraphics[height=.6\textheight,angle=270]{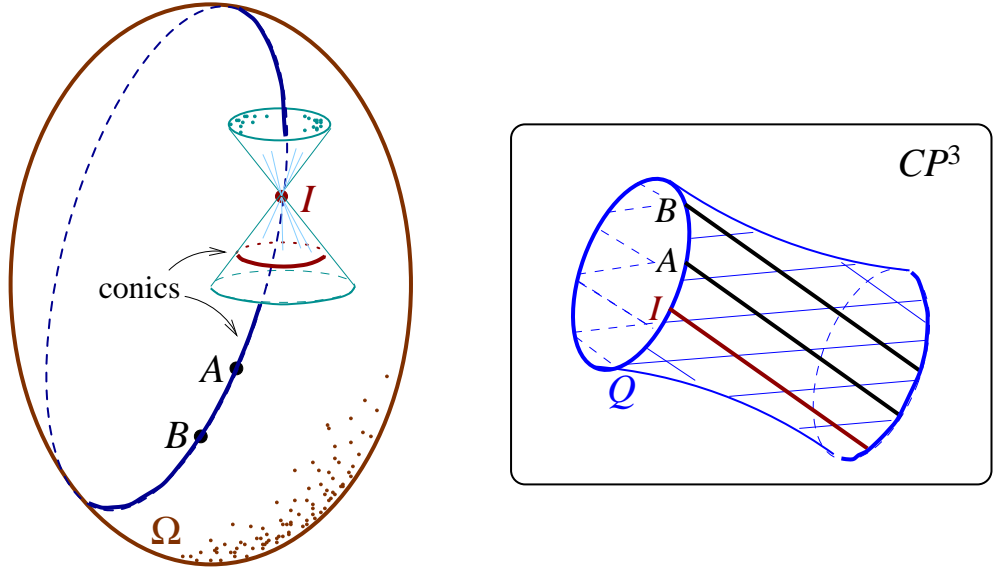}
  \caption{\label{fig:9}{\sl Time-like geodesics in Minkowski
  space.} Each time-like geodesic in a flat four-dimensional
  space-time determines a quadric in projective twistor space. A
  quadric in ${\mathbb P}^3$ has two distinct systems of
  generators. In the case of a time-like geodesic, the line $I$
  belongs to one of the two systems. The corresponding geodesic
  is a conic curve in the Klein quadric $\Omega$, and has the
  property that it passes through the point $I$. The second system
  of generators then corresponds to a polar conic that lies on
  the null cone infinity and does not go through $I$.}
\end{figure}

To pursue this point further, we recall that $\bar{\mathcal H}^4$
has the structure of a quadric $\Omega$ in ${\mathbb P}^5$. More
specifically, for $r=2$ the space of skew rank two twistors is
${\mathbb C}^6$, which is projectively ${\mathbb P}^5$, and
$\bar{\mathcal H}^4$ is the locus defined by the homogeneous
quadratic equation
\begin{eqnarray}
\varepsilon_{\alpha\beta\gamma\delta} X^{\alpha\beta}
X^{\gamma\delta}=0. \label{eq:99}
\end{eqnarray}
Infinity in $\bar{\mathcal H}^4$ can then be defined by the
intersection of $\bar{\mathcal H}^4$ in ${\mathbb P}^5$ with the
projective 4-plane $I^4$ given by the equation
\begin{eqnarray}
\varepsilon_{\alpha\beta\gamma\delta} I^{\alpha\beta}
X^{\gamma\delta}=0.
\end{eqnarray}
If $I^{\alpha\beta}$ is simple, then $I^4$ is tangent to
$\bar{\mathcal H}^4$, and the intersection is a cone---the null
cone at infinity. The geometry of this space plays a crucial role
in determining the properties of time-like geodesics in Minkowski
space, as discussed in Figure~\ref{fig:9}.

On the other hand, if $I^{\alpha\beta}$ is not simple, then the
intersection $I^4\cap\bar{\mathcal H}^4$ is a 3-quadric. The
resulting geometry, if $I^{\alpha\beta}$ is real, is that of
de~Sitter space. The metric on $I^4\cap\bar{\mathcal H}^4$ in this
case is given by
\begin{eqnarray}
\rd s^2 = \frac{\varepsilon_{\alpha\beta\gamma\delta}\rd X^{\alpha
\beta}\rd X^{\gamma\delta}}{(\varepsilon_{\alpha\beta\gamma
\delta} I^{\alpha \beta} X^{\gamma\delta})^2},
\end{eqnarray}
and the parameter $\lambda$ defined by (\ref{eq:105}) has the
interpretation of being the associated cosmological constant. We
note that $\rd s$ is independent of the scale of
$X^{\alpha\beta}$, and has an $I$-weight of $-1$, as is
appropriate for an element of time. The de Sitter group consists
of transformations of ${\mathbb P}^5$ that preserve both the
quadric $\Omega$ and the point $I$.

With the incorporation of some additional structure at infinity,
the entire class of Robertson-Walker cosmological models can be
represented in a similar way \cite{Hurd1,Hurd2,Penrose1,
Penrose6,Penrose8, Penrose9}. The idea can be described briefly as
follows. We start with the complex projective space ${\mathbb
P}^5$ and in it the quadric $\Omega$ defined by (\ref{eq:99}). The
points of space-time are given by a reality structure by requiring
that $\overline{X}^{\alpha\beta}=X^{\alpha\beta}$. Next we
introduce a pencil of 4-planes in ${\mathbb P}^5$ of the form
\begin{eqnarray}
Z_{\alpha\beta}=p P_{\alpha\beta} + q Q_{\alpha\beta} ,
\label{eq:102}
\end{eqnarray}
where $(p,q)\in{\mathbb C}^2-\{0,0\}$. For each such
$Z_{\alpha\beta}$, the intersection of the 4-plane
\begin{eqnarray}
Z_{\alpha\beta}X^{\alpha\beta}=0
\end{eqnarray}
with the real quadric given by
\begin{eqnarray}
\epsilon_{\alpha\beta\gamma\delta}X^{\alpha\beta}X^{\gamma
\delta}=0 \quad {\rm and} \quad \overline{X}^{\alpha\beta}=
X^{\alpha\beta}
\end{eqnarray}
defines a certain subspace of the space-time. For certain choices
of $\{P_{\alpha\beta},Q_{\alpha\beta},p,q\}$ the resulting
subspace can be interpreted (in some cases with the deletion of
certain elements) as a constant-time space-like hypersurface
$\sigma^3$ in the space-time. The corresponding cosmological model
is then obtained by selecting a one-parameter `chronological
family' of such surfaces, and choosing an appropriate conformal
factor for the metric geometry.

In more detail, the construction is as follows. For $k=1$ (a
closed universe) we require the skew twistors $P_{\alpha\beta}$
and $Q_{\alpha\beta}$ to be complex and to satisfy the following
relations:
\begin{eqnarray}
\epsilon^{\alpha\beta\gamma\delta}P_{\alpha\beta} P_{\gamma
\delta}=0, \quad \epsilon^{\alpha\beta\gamma\delta}Q_{\alpha\beta}
Q_{\gamma \delta}=0, \quad {\rm and}\quad
P_{\alpha\beta}=\overline{Q}_{\alpha\beta}.
\end{eqnarray}
We then let $p=\re^{\ri\theta}$, $q=\re^{-\ri\theta}$ for some
$\theta\in[0,2\pi]$. The resulting one-parameter family of
surfaces determined by
\begin{eqnarray}
Z_{\alpha\beta}= \re^{\ri\theta} P_{\alpha\beta} +
\re^{-\ri\theta} \overline{P}_{\alpha\beta} \label{eq:104}
\end{eqnarray}
has the property that each is a 3-sphere. For a $k=-1$ cosmology
we require
\begin{eqnarray}
\epsilon^{\alpha\beta\gamma\delta}P_{\alpha\beta} P_{\gamma
\delta}=0, \quad \epsilon^{\alpha\beta\gamma\delta}Q_{\alpha\beta}
Q_{\gamma \delta}=0,  \quad P_{\alpha\beta}=
\overline{P}_{\alpha\beta}, \quad {\rm and} \quad
Q_{\alpha\beta}=\overline{Q}_{\alpha\beta}.
\end{eqnarray}
In this case the relevant family of 4-planes is given by
(\ref{eq:102}) with $(p,q)\in{\mathbb R}^2-\{0,0\}$, the overall
scale of $Z_{\alpha\beta}$ being unimportant. For $k=0$ we set
\begin{eqnarray}
\epsilon^{\alpha\beta\gamma\delta}P_{\alpha\beta} P_{\gamma
\delta}=0, \quad \epsilon^{\alpha\beta\gamma\delta}Q_{\alpha\beta}
Q_{\gamma \delta}=1,  \quad P_{\alpha\beta}=
\overline{P}_{\alpha\beta}, \quad {\rm and} \quad
Q_{\alpha\beta}=\overline{Q}_{\alpha\beta},
\end{eqnarray}
with $(p,q)$ as above in the $k=-1$ case.

\begin{figure}
  \includegraphics[height=.55\textheight,angle=270]{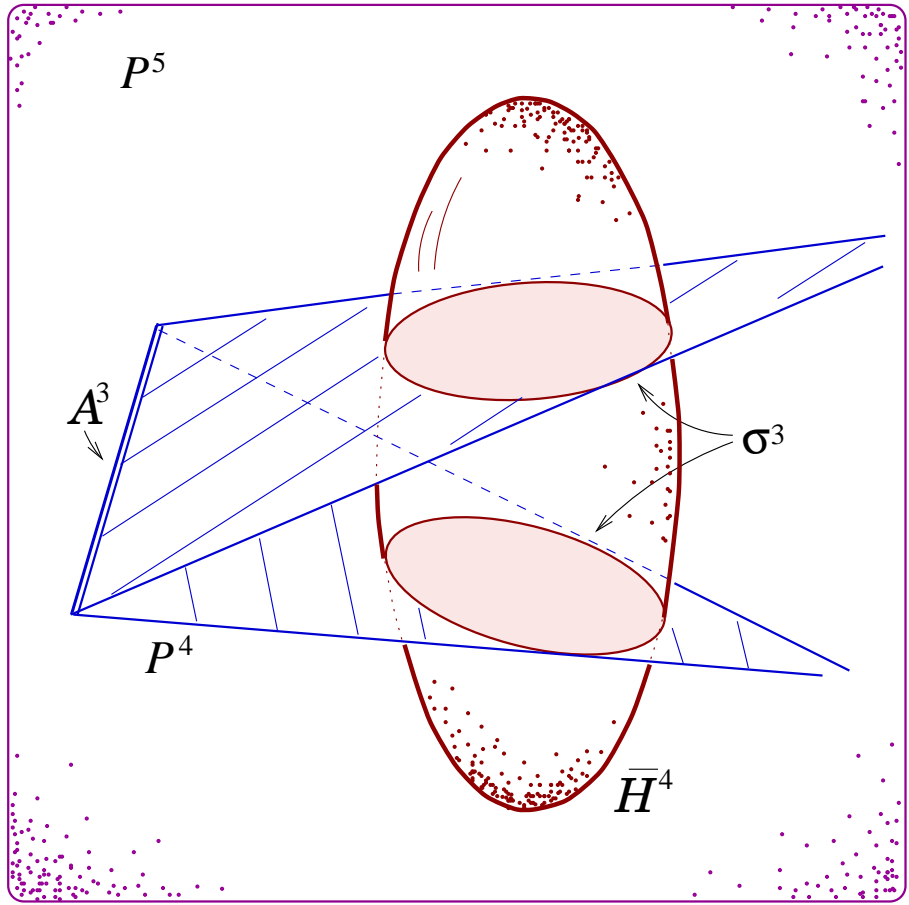}
  \caption{\label{fig:10}{\sl Friedmann-Robertson-Walker
  cosmologies}. In the case $k=1$ depicted here the chronology of
  the FRW cosmology is generated by a pencil of projective
  4-planes hinged on an axis ${\mathcal A}^3$ in ${\mathbb
  P}^5$. The axis is chosen so that it does not impinge on any of
  the real points of the space-time. As a consequence the
  constant-time hypersurfaces in the cosmological model are
  compact.}
\end{figure}

In each case we can consider the `axis' obtained by intersecting
the elements of the given family of 4-planes. In the $k=+1$ case,
the axis itself does not intersect the associated real space-time,
and as a consequence the resulting hypersurfaces of constant time
are topologically 3-spheres (see Figure~\ref{fig:10}). In the
$k=0$ case the axis `touches' the space-time at a point (given by
$q=0$) common to all of the intersection spaces. If we remove this
point (or treat it as a point at infinity), then the resulting
constant-time surfaces are each topologically ${\mathbb R}^3$. In
the $k=-1$ case, the common intersection region is a 2-sphere, and
as a consequence an `open' cosmological model results in this case
as well.

\begin{figure}
  \includegraphics[height=.6\textheight,angle=270]{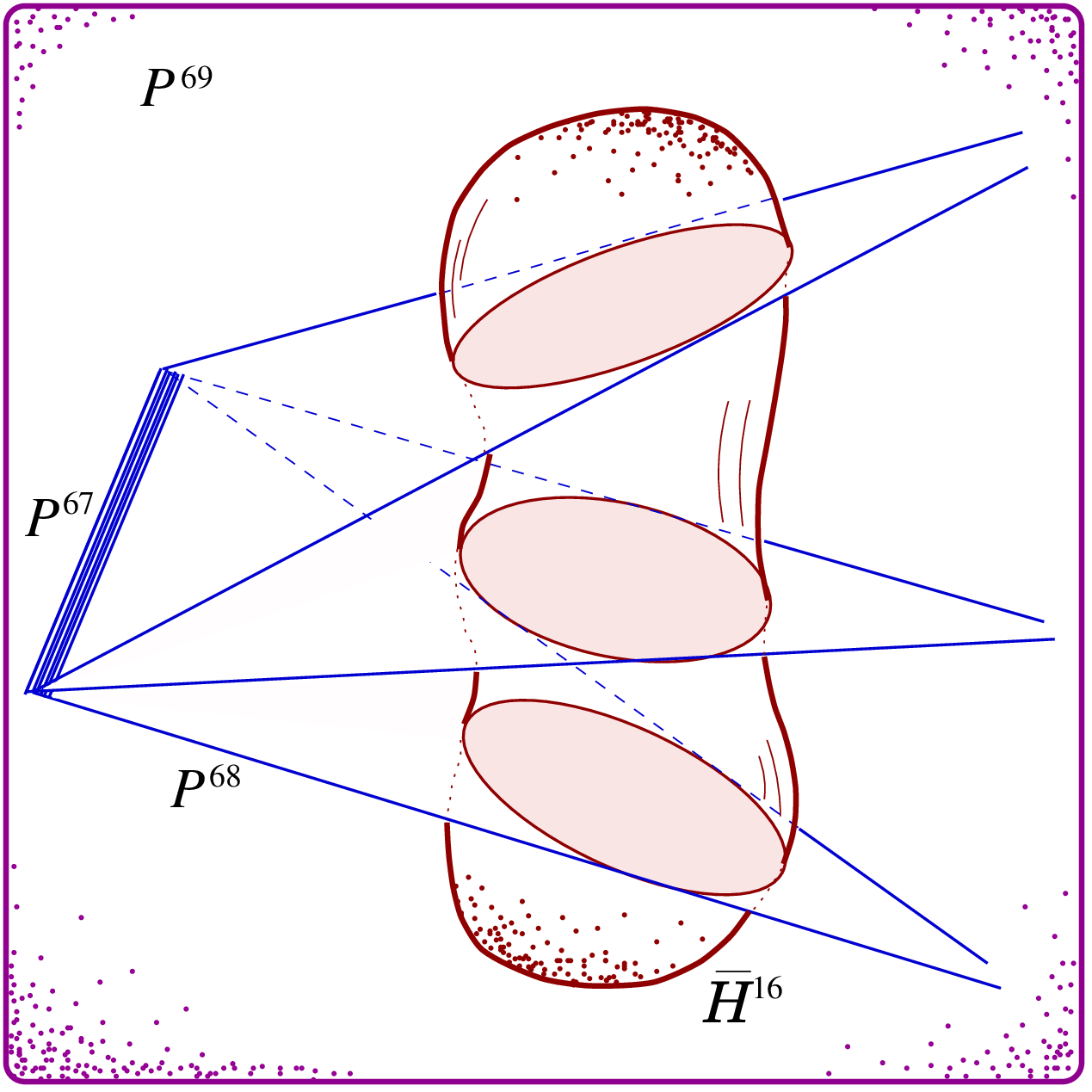}
  \caption{\label{fig:11}{\sl Chronological foliation of an $r=4$
  hypercosmology}. The compactified quantum space-time
  $\bar{\mathcal H}^{16}$ is a real submanifold of the
  complex projective space ${\mathbb P}^{69}$. A pencil of
  ${\mathbb P}^{68}$-hyperplanes hinged on a complex axis
  ${\mathbb P}^{67}$ intersects the space-time in a set of
  hypersurfaces. Depending on the reality structure of the pencil,
  and its real subparameterisation, a variety of different
  possibilities can emerge for the global structure of the
  hypersurface family.}
\end{figure}

Thus we see that the algebraic geometry of twistor theory gives us
an essentially `unified' point of view over the various standard
cosmological models---this approach can be pursued at greater
length, giving rise to a geometrical characterisation of the
different types of situations that can occur, depending, in
particular, on the global structure and topology of the
space-time, on the equation of state of the fluid representation
in the energy tensor, and on the type of cosmological constant (if
any) in the model. Much more detail, along with specific examples
for various choices of the equation of state, can be found in
\cite{Hurd1,Hurd2,Penrose6,Penrose9}.

It is interesting to note that more or less the same state of
affairs prevails in higher dimensions (see Figure~\ref{fig:11} for
an example of a sixteen-dimensional class of `quantum cosmologies'
analogous to the $k=1$ Friedmann-Robertson-Walker models). In
other words, the choice of structure at infinity gives rise to
various possible global structures for the quantum space-time, and
in particular, to a chronometric form that is in general not flat,
thus making $\bar{\mathcal H}^{r^2}$ a cosmological model. In the
case of a standard four-dimensional cosmological model based on
Einstein's theory, the existence of structure at (or `beyond')
infinity has a bearing on the geometry of space-time alone. In the
case of a quantum cosmology, however, the structure at infinity
also has implications for microscopic physics. For instance,
whereas in the four-dimensional de~Sitter cosmology the relevant
structure at infinity contains the `invariant' information of one
dimensional constant (the cosmological constant), in the
higher-dimensional situation there are in general a number of such
constants that may emerge as geometrical invariants of the theory.
Thus within a single geometric framework one has the scope for
introducing structure (or what amounts to the same thing---the
breaking of symmetry) both on a global or cosmological scale, as
well as on the microscopic scales of distance, time, and energy
associated with the phenomenology of elementary particles. One
might say that in these models the structure at infinity is
playing the role of the Higgs fields. One can even envisage the
possibility of explaining, in a purely geometrical language, the
basis of the remarkable coincidences involving various fundamental
constants of nature that have puzzled physicists for many decades.

\section{Weak and strong Hermiticity}

As a prelude to our discussion of the idea of symmetry breaking in
quantum space-time, we digress briefly to review the notions of
weak and strong Hermiticity. This material is relevant to the
origin of unitarity in quantum mechanics. Intuitively speaking, we
observe that when the weak Hermiticity condition is imposed on a
hyperspinor $x^{\,AA'}$ representing a space-time event, then
$x^{\,AA'}$ belongs to the real subspace ${\mathbb R}^{AA'}$. As
such, the hyper-relativistic symmetry of quantum space-time is not
affected by the imposition of this condition. If, however, we
break the hyper-relativistic symmetry by selecting a preferred
time-like direction, then we can speak of a stronger reality
condition whereby an isomorphism is established between the primed
and unprimed hyperspin spaces.

We begin with the weak Hermitian property. Let ${\mathbb S}^{A}$
denote, as before, an $r$-dimensional complex vector space. We
also introduce the associated spaces ${\mathbb S}_{A}$, ${\mathbb
S}^{A'}$, and ${\mathbb S}_{A'}$. In general, there is no natural
isomorphism between ${\mathbb S}^{A'}$ and ${\mathbb S}_{A}$.
Therefore, there is no natural matrix multiplication law or trace
operation defined for elements of ${\mathbb S}^{A}\otimes{\mathbb
S}^{A'}$. Nevertheless, certain matrix operations are well
defined. For example, the determinant of a generic element
$\mu^{AA'}$ is given by
\begin{eqnarray}
\det(\mu) = \frac{1}{r!}\, \varepsilon_{AB\cdots C}\,
\varepsilon_{A'B'\cdots C'}\, \mu^{AA'}\mu^{BB'}\cdots \mu^{CC'}.
\end{eqnarray}
The weak Hermitian property is also well-defined. In particular,
if ${\bar\mu}^{A'A}$ is the complex conjugate of $\mu^{AA'}$, then
we say that $\mu^{AA'}$ is weakly Hermitian if
$\mu^{AA'}={\bar\mu}^{A'A}$. As we have observed, for many
applications, weak Hermiticity suffices.

Now we consider the strong Hermitian property. In some situations
there may exist a natural map ${\mathbb S}^{A'}\to{\mathbb S}_{A}$
defined by the context of the particular problem. Such a map is
called a Hermitian correlation. In this case, the complex
conjugate of an element $\alpha^{A}\in{\mathbb S}^{A}$ determines
an element ${\bar\alpha}_{A}\in{\mathbb S}_{A}$. For any element
$\mu^{A}_{B}\in {\mathbb S}^{A}\otimes{\mathbb S}_{B}$ we define
the operations of determinant, matrix multiplication, and trace in
the usual manner. The determinant is
\begin{eqnarray}
\det(\mu) = \frac{1}{r!}\, \varepsilon_{AB\cdots C}\,
\varepsilon^{PQ\cdots R}\, \mu^{A}_{P}\mu^{B}_{Q} \cdots
\mu^{C}_{R},
\end{eqnarray}
and the Hermitian conjugate of $\mu^{A}_{\ B}$ is
${\bar\mu}_{A}^{\ B}$. The Hermitian correlation is given by the
choice of a preferred element $t_{AA'}\in {\mathbb S}_{A}
\otimes{\mathbb S}_{A'}$. Then we write
\begin{eqnarray}
{\bar\alpha}_{A} = t_{AA'}{\bar\alpha}^{A'},
\end{eqnarray}
where ${\bar\alpha}_{A}$ is now called the complex conjugate of
$\alpha^{A}$. When there is a Hermitian correlation ${\mathbb
S}^{A'} \leftrightarrow{\mathbb S}_{A}$, we call the condition
$\mu^{A}_{\ B}= {\bar\mu}^{A}_{\ B}$ the strong Hermitian
property.

Thus once we break the relativistic invariance by introducing a
preferred element $t_{AA'}$ that determines a Hermitian
correlation, we may carry out specific calculations in that frame.
To gain a better understanding of this, consider an event in
complex Minkowski space defined by its separation $x^{{\bf AA}'}$
from the origin. The complex conjugate of $x^{{\bf AA}'}$ is
${\bar x}^{{\bf AA}'}$, so $x^{{\bf AA}'}$ is real if $x^{{\bf
AA}'}={\bar x}^{{\bf AA}'}$. Let $t^{{\bf AA}'}$ be a fixed
time-like vector satisfying $g_{\rm ab}t^{\rm a}t^{\rm b}=2$. With
respect to this choice of $t^{{\bf AA}'}$, the trace of $x^{{\bf
AA}'}$ is defined by
\begin{eqnarray}
{\rm tr}(x^{{\bf AA}'}) = x^{{\bf AA}'}t_{{\bf AA}'}.
\end{eqnarray}
Once $t^{{\bf AA}'}$ is fixed, we may represent $x^{{\bf AA}'}$ in
terms of Pauli matrices, according to which $x^{{\bf AA}'}$ admits
a matrix representation of the form (\ref{eq:2}). The time
variable is then given by $t=\frac{1}{2} {\rm tr}(x^{{\bf AA}'})$
and the Minkowski metric is given by the determinant (\ref{eq:3}).
We can also define a commutator for a pair of space-time position
vectors $x^{{\bf AA}'}$ and $y^{{\bf AA}'}$ by setting
\begin{eqnarray}
z^{{\bf AA}'} = \left( x^{{\bf AB}'} y^{{\bf BA}'} - y^{{\bf AB}'}
x^{{\bf BA}'} \right) t_{{\bf BB}'}. \label{eq:113}
\end{eqnarray}
The geometrical meaning of (\ref{eq:113}) is that $z^{{\bf AA}'}$
corresponds to the ordinary 3-space cross-product of the
projection of the vectors $x^{{\bf AA}'}$ and $y^{{\bf AA}'}$ onto
the space-like hyperplane orthogonal to $t^{{\bf AA}'}$ that
passes through the origin. Analogously, we can discuss the
`spectral' properties of vectors with respect to a given choice of
$t^{{\bf AA}'}$ and a given choice of origin in space-time. The
two eigenvalues of $x^{{\bf AA}'}$ are then given by $t\pm\sqrt{
x^2+ y^2 +z^2}$. Thus two such vectors $x^{{\bf AA}'}$ and
$y^{{\bf AA}'}$ are isospectral if and only if they lie on a
common sphere about the origin lying in the given space-like
hyperplane. Similar remarks apply to higher-dimensional quantum
space-times.

\section{Symmetry breaking mechanism}

Now we proceed to introduce a natural mechanism for symmetry
breaking that arises in the case of a standard `flat' quantum
space-time endowed with the canonical structures associated with
reality and infinity. We shall make the point in particular that
the breaking of symmetry in quantum space-time is intimately
linked to the notion of quantum entanglement. According to this
point of view the introduction of symmetry-breaking in the early
stages of the universe can be understood as a phase transition, or
a sequence of phase transitions, the ultimate consequence of which
is an approximate disentanglement of a four-dimensional
`classical' space-time.

In practical terms the breaking of symmetry is represented in our
framework by an `index decomposition'. In particular, if the
dimension $r$ of the hyperspin space is not a prime number, then a
natural method of breaking the symmetry arises by consideration of
the decomposition of $r$ into factors. The specific essential
assumption that we shall make at this juncture will be that the
dimension of the hyperspin space ${\mathbb S}^{A}$ is \emph{even}.
Then we write $r=2n$, where $n=1,2,\ldots$, and set
\begin{eqnarray}
{\mathbb S}^{A}={\mathbb S}^{{\bf A}i},
\end{eqnarray}
where ${\bf A}$ is a standard two-component spinor index, and $i$
will be called an `internal' index $(i=1,2, \ldots,n)$. Thus we
can write ${\mathbb S}^{{\bf A}i}={\mathbb S}^{{\bf A}}
\otimes{\mathbb H}^{i}$, where ${\mathbb S}^{{\bf A}}$ is a
standard spin space of dimension two, and ${\mathbb H}^{i}$ is a
complex vector space of dimension $n$. The breaking of the
symmetry then amounts to the fact that we can identify the
hyperspin space with the tensor product of these two spaces.

We shall assume, moreover, that ${\mathbb H}^{i}$ is endowed with
a strong Hermitian structure, i.e. we shall assume that there is a
canonical anti-linear isomorphism between the complex conjugate of
the internal space ${\mathbb H}^{i}$ and the dual space ${\mathbb
H}_{i}$. If $\psi^i\in{\mathbb H}^{i}$, then we write
${\bar\psi}_i$ for the complex conjugate of $\psi^i$, where
${\bar\psi}_i\in{\mathbb H}_{i}$. We see therefore that ${\mathbb
H}^{i}$ is a complex Hilbert space---and indeed although here we
consider for technical simplicity the case for which $n$ is
finite, one should have in mind also the general infinite
dimensional situation. For the other hyperspin spaces we write
\begin{eqnarray}
{\mathbb S}_{A}={\mathbb S}_{{\bf A}i},\quad {\mathbb
S}^{A'}={\mathbb S}^{{\bf A}'}_{\ \ i},\quad {\rm and}\quad
{\mathbb S}_{A'}={\mathbb S}_{{\bf A}'}^{\ \ i},
\end{eqnarray}
respectively. These equivalences preserve the duality between
${\mathbb S}^{A}$ and ${\mathbb S}_{A}$, and between ${\mathbb
S}^{A'}$ and ${\mathbb S}_{A'}$; and at the same time are
consistent with the complex conjugation relations between
${\mathbb S}^{A}$ and ${\mathbb S}^{A'}$, and between ${\mathbb
S}_{A}$ and ${\mathbb S}_{A'}$. Hence if $\alpha^{{\bf A}i}
\in{\mathbb S}^{{\bf A}i}$ then under complex conjugation we have
$\alpha^{{\bf A}i}\to {\bar\alpha}^{{\bf A}'}_{\ \ i}$, and if
$\beta_{{\bf A}i}\in{\mathbb S}_{{\bf A}i}$ then $\beta_{{\bf
A}i}\to{\bar\beta}_{{\bf A}'}^{\ \ i}$.

In the case of a quantum space-time vector $r^{\,AA'}$ we have a
corresponding structure induced by the identification
\begin{eqnarray}
r^{\,AA'}=r^{{\bf AA}'i}_{\ \ \ \ \ j}.
\end{eqnarray}
When the quantum space-time vector is real, the weak Hermitian
structure on $r^{\,AA'}$ is manifested in the form of a standard
weak Hermitian structure on the two-component spinor index pair,
together with a strong Hermitian structure on the internal index
pair. In other words, the Hermitian condition on the space-time
vector $r^{\,AA'}$ is given by
\begin{eqnarray}
{\bar r}^{{\bf A}'{\bf A}\ \ i}_{\ \ \ \ \ j} = r^{{\bf AA}'i}_{\
\ \ \ \ j} .
\end{eqnarray}

One consequence of these relations is that we can interpret each
point in quantum space-time as being a space-time valued operator.
The ordinary classical space-time then `sits' inside the quantum
space-time in a canonical manner---namely, as the locus of those
points of quantum space-time that factorise into the product of a
space-time point $x^{{\bf AA}'}$ and the identity operator on the
internal space:
\begin{eqnarray}
x^{{\bf AA}'i}_{\ \ \ \ \ j} = x^{{\bf AA}'} \delta^{i}_{\ j}.
\label{eq:11.4}
\end{eqnarray}
Thus, in situations where special relativity is a satisfactory
theory, we regard the relevant events as taking place on or in the
immediate neighbourhood of this embedding of Minkowski space in
${\mathcal H}^{4n^2}$.

This picture can be presented in more geometric terms as follows.
The hypertwistor space ${\mathbb P}^{2r-1}$ in the case $r=2n$
admits a Segr\'e embedding of the form
\begin{eqnarray}
{\mathbb P}^3 \times {\mathbb P}^{n-1}\subset{\mathbb P}^{4n-1}.
\end{eqnarray}
Many such embeddings are possible, though they are all equivalent
to one another under the action of the overall symmetry group
$U(2n,2n)$. If the symmetry is broken and one such embedding is
selected out, then following the conventions discussed earlier we
can introduce homogeneous coordinates and write $Z^{\alpha i}$ for
the hypertwistor. Here the Greek letter ${\alpha}$ denotes an
ordinary twistor index $({\alpha}=0,1,2,3)$ and $i$ denotes an
internal index $(i=1,2,\ldots,n)$. These two indices, when clumped
together, constitutes a hypertwister index. The Segr\'e embedding
consists of those points in ${\mathbb P}^{4n-1}$ for which we have
a product decomposition of the associated hypertwistor given by
\begin{eqnarray}
Z^{\alpha i}= Z^{\alpha}\psi^i. \label{eq:120}
\end{eqnarray}

The idea of symmetry breaking that we are putting forward here is
related to the notion of disentanglement in standard quantum
mechanics (cf. Gibbons 1992; Brody \& Hughston 2001). That is to
say, at the unified level the degrees of freedom associated with
space-time symmetry are quantum mechanically entangled with the
internal degrees of freedom associated with microscopic physics.
The phenomena responsible for the breakdown of symmetry are thus
analogous to the mechanisms of decoherence through which quantum
entanglements are gradually diminished. Some readers may raise the
objection that surely it is impossible to unify the unitary
symmetries of elementary particle phenomenology with the
symmetries of space-time (cf., e.g., \cite{Coleman}). It should be
noted, however, that our approach is not to attempt to embed a
relativistic symmetry group in a higher-dimensional unitary group,
but rather to embed the unitary group in a higher-dimensional
relativistic symmetry group. Our methodology is consistent with
the point of view put forward by Penrose that for a coherent
unification of general relativity and quantum mechanics, the rules
of quantum theory must undergo `profound
modification'~\cite{Penrose7}.

The compactified complexified quantum space-time ${\mathcal C}\!
\bar{\mathcal H}^{4n^2}$ can be regarded as the aggregate of
projective $(2n-1)$-planes in ${\mathbb P}^{4n-1}$. Now
generically a ${\mathbb P}^{2n-1}$ in ${\mathbb P}^{4n-1}$ will
not intersect the Segr\'e variety
\begin{eqnarray}
{\mathcal G}^{n+2} = {\mathbb P}^{3}\times{\mathbb P}^{n-1}.
\end{eqnarray}
Such a generic $(2n-1)$-plane corresponds to a generic point in
${\mathcal C}\!\bar{\mathcal H}^{4n^2}$. The $(2n-1)$-planes that
correspond to the points of compactified complexified Minkowski
space can be constructed as follows. For each line $L$ in
${\mathbb P}^{3}$ we consider the subvariety ${\mathcal
G}_L^n\subset{\mathcal G}^{n+2}$ where ${\mathcal G}_L^n ={\mathbb
P}_L^1\times{\mathbb P}^{n-1}$. For any algebraic variety
$V^j\subset{\mathbb P}^l$ $(j\leq l-1)$ we define the \emph{span}
of $V^j$ to be the projective plane spanned by the points of
$V^j$. We say a point $X$ in the ambient space ${\mathbb P}^l$
lies in the span of the variety $V^j$ if and only if there exist
$m$ points in $V^j$ for some $m\geq2$ with the property that $X$
lies in the $(m-1)$-plane spanned by those $m$ points. The
dimension $k$ of the span of $V^j$ satisfies $j\leq k\leq l$;
however, the value of $k$ depends on the geometry of $V^j$.

The linear span of the points in ${\mathcal G}_L^n$, for any given
$L$, is a $(2n-1)$-plane. This is the ${\mathbb P}_L^{2n-1}$ in
${\mathbb P}^{4n-1}$ that represents the point in $\bar{\mathcal
H}^{4n^2}$ corresponding to the line $L$ in ${\mathbb P}^3$. The
aggregate of such special $(2n-1)$-planes, defined by their
intersection properties with the Segr\'e variety ${\mathcal
G}^{n+2}$, constitutes a submanifold of ${\mathcal C}\!
\bar{\mathcal H}^{4n^2}$, and this submanifold is compactified
complexified Minkowski space ${\mathcal C}\!\bar{\mathcal H}^{4}$.
Thus we see that once the symmetry of quantum space-time
${\mathcal C}\!{\mathcal H}^{4n^2}$ has been broken in the
particular way we have discussed, then ordinary Minkowski space
can be identified as a submanifold.

Let us now consider the implications of our symmetry breaking
mechanism for fields defined on quantum space-time. As an example,
let $\phi(x^{\,AA'})$ be a scalar field on quantum space-time.
After we break the symmetry by writing $x^{\,AA'}= x^{{\bf
AA}'i}_{\ \ \ \ \ j}$, we consider a Taylor expansion of the field
around the embedded Minkowski space-time. Specifically, for such
an expansion we have
\begin{eqnarray}
\phi(x^{\,AA'}) = \phi^{(0)}(x^{{\bf AA}'}) + \phi^{(1)\ i}_{{\bf
AA}'\ j}(x^{\,{\bf AA}'})\left( x^{{\bf AA}'j}_{\ \ \ \ \ i} -
x^{{\bf AA}'}\delta^j_{\ i} \right) + \cdots,
\end{eqnarray}
where
\begin{eqnarray}
\phi^{(0)}(x^{{\bf AA}'}) = \phi(x^{\,AA'})\Big|_{x^{\,AA'}
=x^{{\bf AA}'} \delta^{j}_{\ i}}\ ,
\end{eqnarray}
and
\begin{eqnarray}
\phi^{(1)\ i}_{{\bf AA}'\ j} (x^{\,{\bf AA}'}) = \left.
\frac{\partial \phi}{
\partial x^{{\bf AA}'j}_{\ \ \ \ \ i}} \right|_{x^{{\bf AA}'j}_{\
\ \ \ \ i} =x^{{\bf AA}'} \delta^{j}_{\ i}}.
\end{eqnarray}
Therefore, the order zero term has the character of a classical
field on Minkowski space, and the first order term can be
interpreted as a `multiplet' of fields, transforming according to
the adjoint representation of the internal symmetry group $U(n)$.

\begin{figure}
  \includegraphics[height=.62\textheight,angle=270]{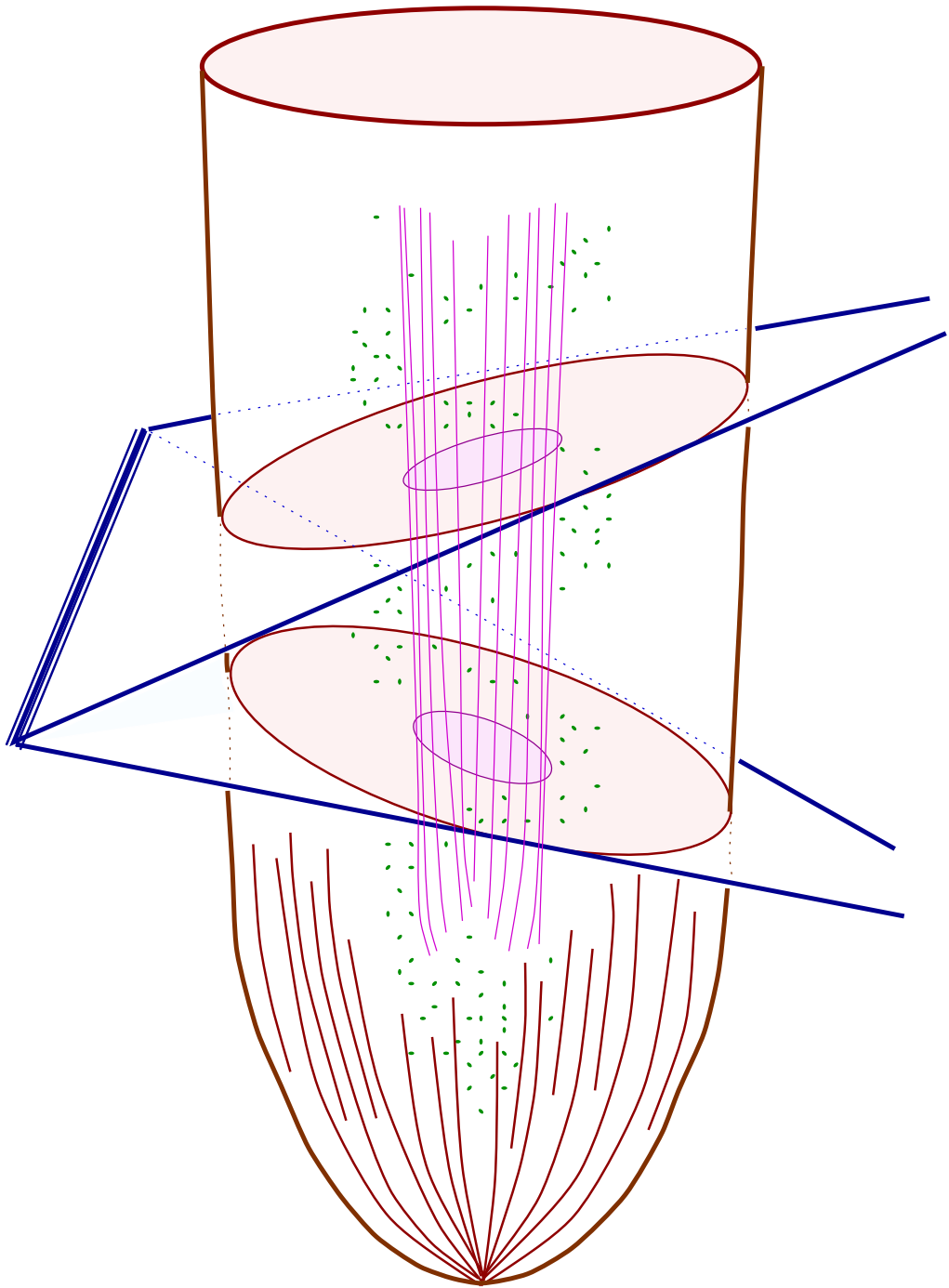}
  \caption{\label{fig:12}{\sl Symmetry breaking and matter
  formation in a hypertwistorial quantum cosmology}. In its
  earliest stages the universe is highly
  symmetrical. Eventually, symmetry is broken, and a
  four-dimensional `classical' cosmology freezes out, becoming
  largely disentangled from the reminder of the quantum
  space-time. The formation of matter during the disentanglement
  process may not be confined to the four-dimensional subspace. If
  such a scenario prevails, the bulk of this `dark material' is
  likely to remain outside the four-dimensional subspace, though
  nevertheless having an impact on its dynamics.}
\end{figure}

In this connection we note that the symmetry breaking mechanism
that we have proposed here has yet another
representation---namely, the expression of a hypertwistor as a
multi-twistor system, i.e. as a multiplet of Penrose twistors. The
physical and geometrical characteristics of such $n$-twistor
systems have been analysed at great length by a number of authors
(see, e.g., \cite{Hughston2,Hughston3,Penrose3,Penrose4,
Penrose5,Sparling,Tod1} and references cited therein), and it is
interesting therefore to see the direct link with hypertwistor
theory and quantum space-time geometry. It is fitting also to make
a tribute here to the work of Zoltan Perj\'es, whose extensive
contributions to relativity theory include, in particular, a
number of important studies concerning the properties of
$n$-twistor systems and their
symmetries~\cite{Lukacs,Perjes1,Perjes2, Perjes3,Perjes4,Perjes5,
Perjes6,Perjes7,Perjes8,Perjes9, Perjes10,Perjes11,Tod2}.

It is tempting to speculate that even in a more dynamic context
some version of the symmetry breaking mechanism provided here will
manifest itself. In this picture we would envisage the earliest
stages of the universe as being highly symmetrical, rather in the
spirit of Penrose's Weyl curvature hypothesis~\cite{Penrose11,
Penrose12}, with no appreciable distinction between the
conventional space-time degrees of freedom and the internal
degrees of freedom associated with quantum theory. Nevertheless,
the causal geometry of the universe remains well defined, and it
is interesting to ask whether there might be some scenario within
the rather rich causal structure of a quantum space-time that
would allow us to account for the so-called `horizon problem'. In
any event, once symmetry breaking takes place---and this may
happen in stages, corresponding to a successive factorisation of
the underlying hypertwistor space---then it makes sense to think
of ordinary four-dimensional space-time as becoming more or less
disentangled from the rest of the universe, and behaving in a way
that is to some extent autonomous. Nonetheless, we might
reasonably expect its global dynamics, on a cosmological scale, to
be affected by the distribution of mass and energy elsewhere in
the quantum space-time as well (see, e.g., Figure~\ref{fig:12}).

\section{Emergence of quantum probability}

The embedding of Minkowski space in the quantum space-time
${\mathcal H}^{4n^2}$ given by (\ref{eq:11.4}) implies a
corresponding embedding of the Poincar\'e group in the
hyper-Poincar\'e group. This can be seen as follows. The standard
Poincar\'e group in ${\mathcal H}^{4}$ consists of transformations
of the form
\begin{eqnarray}
x^{{\bf AA}'}\longrightarrow\, \lambda^{\bf A}_{\bf B}\,
\bar{\lambda}^{{\bf A}'}_{{\bf B}'}\, x^{{\bf BB}'} + \beta^{{\bf
AA}'}, \label{eq:12.1}
\end{eqnarray}
and the hyper-Poincar\'e transformations in ${\mathcal H}^{4n^2}$
are of the form
\begin{eqnarray}
x^{{\bf AA}'i}_{\ \ \ \ \ j}\longrightarrow\, \lambda^{\bf A}_{\bf
B}\, \bar{\lambda}^{{\bf A}'}_{{\bf B}'}\, x^{{\bf BB}'i}_{\ \ \ \
\ j} + \beta^{{\bf AA}'}\delta^i_{\,j}. \label{eq:12.2}
\end{eqnarray}
With the identification $A={\bf A}i$, the general hyper-Poincar\'e
transformation in the broken symmetry phase can be expressed in
the form
\begin{eqnarray}
x^{{\bf AA}'i}_{\ \ \ \ \ j}\longrightarrow\, L^{{\bf A}i}_{{\bf
B}k}\, \bar{L}^{{\bf A}'l}_{{\bf B}'j}\, x^{{\bf BB}'k}_{\ \ \ \ \
l} + \beta^{{\bf AA}'i}_{\ \ \ \ \ j}. \label{eq:12.3}
\end{eqnarray}
Thus the embedding of the Poincar\'e group as a subgroup of the
hyper-Poincar\'e group is given by
\begin{eqnarray}
L^{{\bf A}i}_{{\bf B}j} = \lambda^{\bf A}_{\bf B}\delta^i_{\ j}
\quad {\rm and} \quad \beta^{{\bf AA}'i}_{\ \ \ \ \ j} =
\beta^{{\bf AA}'} \delta^i_{\,j}. \label{eq:12.4}
\end{eqnarray}

Bearing this in mind, we now construct a class of maps from the
general even-dimensional quantum space-time ${\mathcal H}^{4n^2}$
to Minkowski space ${\mathcal H}^{4}$. It turns out~\cite{Brody2}
that under rather general physical assumptions such maps are
necessarily of the form
\begin{eqnarray}
x^{{\bf AA}'i}_{\ \ \ \ \ j}\longrightarrow\, x^{{\bf AA}'} =
\rho^j_i\,x^{{\bf AA}'i}_{\ \ \ \ \ j}, \label{eq:12.5}
\end{eqnarray}
where $\rho^j_i$ is a density matrix. As usual, by a density
matrix we mean a positive semi-definite Hermitian matrix with unit
trace. Thus the maps arising here can be regarded as quantum
expectations.

In particular, let $\rho:\ {\mathcal H}^{4n^2}\to {\mathcal H}^4$
satisfy the following conditions: {\rm (i)} $\rho$ is linear and
maps the origin of ${\mathcal H}^{4n^2}$ to the origin of
${\mathcal H}^{4}$; {\rm (ii)} $\rho$ is Poincar\'e invariant; and
{\rm (iii)} $\rho$ preserves causal relations. Then $\rho$ is
given by a density matrix on the internal space. \label{theo:2}

The general linear map from ${\mathcal H}^{4n^2}$ to ${\mathcal
H}^{4}$ preserving the origin is given by
\begin{eqnarray}
x^{{\bf AA}'i}_{\ \ \ \ \ j}\longrightarrow\, x^{{\bf AA}'} =
\rho^{{\bf AA}'j}_{{\bf BB}'i}\, x^{{\bf BB}'i}_{\ \ \ \ \ j},
\label{eq:12.6}
\end{eqnarray}
where $\rho^{{\bf AA}'j}_{{\bf BB}'i}$ is weakly Hermitian. Now
suppose that we subject ${\mathcal H}^{4n^2}$ to a Poincar\'e
transformation of the form {\rm (}\ref{eq:12.2}{\rm )}, and
require the corresponding transformation of ${\mathcal H}^4$
should be of the form {\rm (}\ref{eq:12.1}{\rm )}. If $\rho$
satisfies these conditions then we shall say that the map $\rho$
is Poincar\'e invariant. Clearly, Poincar\'e invariance holds if
and only if
\begin{eqnarray}
\rho^{{\bf AA}'j}_{{\bf BB}'i}\left( \lambda^{\bf B}_{\bf C}\,
\bar{\lambda}^{{\bf B}'}_{{\bf C}'}\,x^{{\bf CC}'i}_{\ \ \ \ \ j}
+ b^{{\bf BB}'}\delta^i_{\,j} \right) =  \lambda^{\bf A}_{\bf B}\,
\bar{\lambda}^{{\bf A}'}_{{\bf B}'}\left( \rho^{{\bf BB}'j}_{{\bf
CC}'i} \,x^{{\bf CC}'i}_{\ \ \ \ \ j}\right) + \beta^{{\bf AA}'}
\label{eq:12.7}
\end{eqnarray}
for all $\lambda^{\bf A}_{\bf B}\in SL(2n,{\mathbb C})$, for all
$\beta^{{\bf AA}'}\in{\mathbb S}^{{\bf AA}'}$, and for all
$x^{{\bf AA}'i}_{\ \ \ \ \ j}\in{\mathcal H}^{4n^2}$. Thus we have
\begin{eqnarray}
\rho^{{\bf AA}'j}_{{\bf BB}'i}\, \lambda^{\bf B}_{\bf C}\,
\bar{\lambda}^{{\bf B}'}_{{\bf C}'} = \lambda^{\bf A}_{\bf B}\,
\bar{\lambda}^{{\bf A}'}_{{\bf B}'}\, \rho^{{\bf BB}'j}_{{\bf
CC}'i}\label{eq:12.8}
\end{eqnarray}
for all $\lambda^{\bf A}_{\bf B}$, and
\begin{eqnarray}
\rho^{{\bf AA}'j}_{{\bf BB}'i}\,\delta^{i}_{\,j}\, \beta^{{\bf
BB}'} = \beta^{{\bf AA}'} \label{eq:12.9}
\end{eqnarray}
for all $\beta^{{\bf AA}'}$. Equation (\ref{eq:12.8}) implies that
$\rho$ is of the form
\begin{eqnarray}
\rho^{{\bf AA}'j}_{{\bf BB}'i} = \delta^{\bf A}_{\,\bf B}\,
\delta^{{\bf A}'}_{\,{\bf B}'}\,\rho^{j}_{i} \label{eq:12.10}
\end{eqnarray}
for some $\rho^j_i$. Then {\rm (}\ref{eq:12.9}{\rm )} implies that
$\rho$ must satisfy the trace condition $\rho^i_i=1$. Finally we
require that if $x^{{\bf AA}'i}_{\ \ \ \ \ j}$ and $y^{{\bf
AA}'i}_{\ \ \ \ \ j}$ are quantum space-time points with the
property that the interval
\begin{eqnarray}
r^{{\bf AA}'i}_{\ \ \ \ \ j}=x^{{\bf AA}'i}_{\ \ \ \ \ j}-y^{{\bf
AA}'i}_{\ \ \ \ \ j}
\end{eqnarray}
is future-pointing then $r^{{\bf AA}'}=x^{{\bf AA}'}-y^{{\bf
AA}'}$ is also future pointing, where
\begin{eqnarray}
r^{{\bf AA}'}=\rho^j_i\,r^{{\bf AA}'i}_{\ \ \ \ \ j}.
\end{eqnarray}
This is the requirement that $\rho$ should be a `causal' map. This
condition implies that $\rho$ must be positive semi-definite. In
particular, if $r^{{\bf AA}'i}_{\ \ \ \ \ j}$ is future-pointing
then it must be of the form
\begin{eqnarray}
r^{{\bf AA}'i}_{\ \ \ \ \ j} = \xi^{{\bf A}i} \bar{\xi}^{{\bf
A}'}_j + \eta^{{\bf A}i} \bar{\eta}^{{\bf A}'}_j + \cdots
.\label{eq:12.11}
\end{eqnarray}
Consider therefore the case for which $r^{{\bf AA}'i}_{\ \ \ \ \
j}$ is null. Then we require that the expression $\xi^{{\bf A}i}
\bar{\xi}^{{\bf A}'}_j\rho^j_i$ should be future-pointing (or
vanish) for any choice of $\xi^{{\bf A}i}$. In particular, we
require that the vector $\xi^{{\bf A}i} \bar{\alpha}^{{\bf
A}'}_j\rho^j_i$ should be future-pointing if $\xi^{{\bf A}i}$ is
of the form
\begin{eqnarray}
\xi^{{\bf A}i}=\alpha^{\bf A} \psi^i
\end{eqnarray}
for any choice of $\xi^{\bf A}$ and $\psi^i$. This means that the
inequality
\begin{eqnarray}
\rho^j_i \psi^i\bar{\psi}_j \geq0
\end{eqnarray}
holds for all $\psi^i$, which shows that $\rho^j_i$ is positive
semi-definite. Since we have shown that the trace of $\rho^j_i$ is
unity, it follows that $\rho^j_i$ is a density matrix.

This result shows how the causal structure of quantum space-time
is linked with the probabilistic structure of quantum mechanics.
The concept of a quantum state emerges when we ask for consistent
ways of `averaging' over the geometry of quantum space-time in
order to obtain a reduced description of physical phenomena in
terms of the geometry of Minkowski space. We see that a
probabilistic interpretation of the map from a general quantum
space-time to Minkowski space arises as a consequence of
elementary causality requirements. We can thus view the space-time
events in ${\mathcal H}^{4n^2}$ as representing quantum
observables, the expectations of which correspond to points of
${\mathcal H}^{4}$.


\begin{theacknowledgments}
DCB gratefully acknowledges financial support from The Royal
Society. The work described here is based, in part, on ideas and
suggestions arising in discussions with E.~J.~Brody. The authors
are grateful to participants at the XIXth Max Born Symposium,
Institute of Theoretical Physics, Wroclaw, Poland, for helpful
comments.
\end{theacknowledgments}

\bibliographystyle{aipproc}   

\end{document}